\documentclass[12pt,a4paper]{article}

\usepackage[english]{babel}

\usepackage[a4paper,top=2cm,bottom=2cm,left=2.5cm,right=2.5cm,marginparwidth=1.75cm]{geometry}

\usepackage[backend=biber, style=ieee, citestyle=numeric-comp, sorting=none]{biblatex}
\usepackage{amsmath}
\usepackage{amssymb}
\usepackage{physics}
\usepackage{bm}
\usepackage{graphicx}
\usepackage[colorlinks=true, allcolors=blue]{hyperref}
\usepackage{hyperref}
\usepackage[title]{appendix}
\usepackage{mathrsfs}
\usepackage{amsfonts}
\usepackage{booktabs} 
\usepackage{threeparttable} 
\usepackage{algorithm}
\usepackage{algorithmicx}
\usepackage{algpseudocode}
\usepackage{listings}
\usepackage{enumitem}
\usepackage{chngcntr}
\usepackage{booktabs}
\usepackage{lipsum}
\usepackage{subcaption}
\usepackage{authblk}
\usepackage[T1]{fontenc}    
\usepackage{csquotes}       
\usepackage{diagbox}
\usepackage{comment}
\usepackage{url}
\usepackage{helvet}  
\usepackage{lmodern}  

\usepackage{setspace}
\usepackage{titlesec}
\titleformat{\section} 
{\normalfont\Large\bfseries}{\thesection.}{1em}{}

\usepackage{float}   
\usepackage{caption}


\begin{document}


\begin{center}
  {\Huge MAPS: A Novel Multi-Axial Projective Sphere for Geometrically Visualizing Higher $d$-Valued Quantum State-Space of Qudits}
  \\[5ex]

  {\large Ali Al-Bayaty} \\
  {\small
    Department of Electrical and Computer Engineering, Portland State University \\ Portland, Oregon 97201, USA \\
    albayaty@pdx.edu \\
    \url{https://orcid.org/0000-0003-2719-0759}
  } \\ [1.2em]

\end{center}

\begin{abstract}
  Visualizing the $d$-valued quantum state-space of quantum systems serves as a foundational pillar for the scientific research and practical applications in quantum computing and information science, where $d \geq 2$. The 2-valued quantum states of a qubit are elegantly visualized on the three-dimensional Bloch sphere. In contrast, expanding this geometrical $S^2$ paradigm to visualize higher $d$-valued quantum states of a qudit ($d \geq 3$), e.g., a qutrit ($d = 3$), ququadit ($d = 4$), and quintit ($d = 5$), leads to severe structural and topological complexities. This paper introduces a new generalized three-dimensional $S^2$ framework to effectively visualize higher $d$-valued quantum states of a qudit, in the aspects of ease of illustration, structural simplicity, and natural representation for researchers and engineers. We called this new $S^2$ framework the ``multi-axial projective sphere (MAPS)'', which consists of $n$ projectional intersecting spatial axes, where $0 \leq n \leq d-1$. We also propose a group of novel $d$-valued phase axial-based gates, to swivel and shift $d$-valued quantum states of a qudit along these $n$ axes. Our generalized $S^2$ framework could be used for visualizing high-dimensional data for practical applications, e.g., machine learning, quantum machine learning, and quantum chemistry, where every axis of the MAPS represents a single feature of such data with its corresponding distinct values.
\end{abstract}

\noindent
\textbf{\textit{Keywords}:} Quantum state-space, $d$-valued quantum states, qutrit, ququadit, quintit, qudit, visualization framework, Bloch sphere, phase gates, swivel gates, shift gates

\section{Introduction}
The visualization of state-space of quantum systems represents an essential connection between both abstract functional analysis and physical intuition. In quantum mechanics, the geometrical interpretations of quantum pure (basis) and mixed (superimposed) states have driven fundamental discoveries in quantum computing and information theory. The most well-known geometrical interpretation is the ``Bloch sphere'' \cite{bloch1946, bloch1946nuclear, bloch1953}, as a visualization framework of unit $S^2$ sphere. The Bloch sphere encapsulates the pure and mixed states of a 2-valued quantum information, i.e., a quantum bit (a qubit), within a three-dimensional binary Euclidean ball $\left(\mathbb{B}^3 \subset \mathbb{R}^3\right)$ of three intersecting axes (X, Y, and Z). Mathematically, such an interpretation influences the exceptional isomorphism between the Lie groups $SU(2)$ and $SO(3)$ \cite{silverman2022, chevalley2018, duistermaat2012}, allowing any $2 \times 2$ density matrix $\rho$ to be mapped injectively to a real polarization vector $r = (x, y, z)^T$. Notice that, in Lie groups, the $SU(2)$ stands for ``Special Unitary'' group of degree 2, which represents the set of $2 \times 2$ unitary matrices with a determinant of $1$, while the $SO(3)$ stands for ``Special Orthogonal'' group of degree 3, which represents the set of $3 \times 3$ orthogonal matrices with a determinant of $1$. In physics, the $SU(2)$ is used to describe spin and isospin, while the $SO(3)$ corresponds to all possible three-dimensional spatial rotations.

However, as quantum computing systems evolve toward the utilization of higher-dimensional quantum state-space of a $d$-valued quantum information, which is the so-called ``qudit'', the geometrical interpretation simplicity of the Bloch sphere collapses, where $d \geq 3$. Notice that $d=3$ for a ternary quantum information, which is the so-called ``qutrit'', $d=4$ for a quaternary quantum information, which is the so-called ``ququadit'', $d=5$ for a quinary quantum information, which is the so-called ``quintit'', and so on.

In general, the quantum state-space of a qudit has an operational dimensionality of $d^2-1$, where $d \geq 3$. For instance, the quantum state-space of a qutrit resides within an 8-dimensional manifold embedded in $\mathbb{R}^8$, governed by the Lie group $SU(3)$. Every vector for a qubit on the Bloch sphere boundary represents a pure 2-valued state, while the quantum state-space of a qudit is bounded by a complicated-shaped convex body known as the ``generalized Bloch body $\left(\Omega_d\right)$'' \cite{cobanera2018, maciejko2022, dobardvzic2015}. Notice that the partial boundary of $\Omega_d$ contains nested layers of varying geometric and algebraic dimensions, where only a tiny fraction of its surface area corresponds to $d$-valued pure states.

The goal of this paper is to introduce and formalize a new generalized three-dimensional $S^2$ framework to visualize the quantum state-space of a qudit, in the aspects of ease of illustration, structural simplicity, and natural representation of $d$-valued quantum states for engineers and researchers, without using any complicated geometries and topological structures. We called this new visualization $S^2$ framework the ``multi-axial projective sphere (MAPS)''. In general, the MAPS visualizes any higher $d$-valued quantum state-space of a qudit for $d \geq 3$, since the Bloch sphere is an adequate $S^2$ framework for $d=2$.

For the Bloch sphere, the global and relative (local) phases of the quantum states for a qubit are not directly visualized along all three spatial axes. For this reason, additional mathematical formulations are required to describe the complete 2-valued quantum state-space of a qubit. In contrast, our introduced MAPS is geometrically constructed using $n$ projectional intersecting spatial axes, where $0 \leq n \leq d-1$. Every spatial axis is mapped to one $d$-valued quantum state with its corresponding set of relative phases, and the $|0\rangle$-axis of the MAPS always indicates the global phase for the overall $d$-valued quantum state-space. Therefore, as an advantageous feature compared to the Bloch sphere, the MAPS can directly visualize the global and relative phases to describe the complete $d$-valued quantum state-space of a qudit, without requiring any mathematical formulations.

In this paper, we also propose a group of novel $d$-valued phase axial-based gates, to swivel (rotate) and shift (scale) the $d$-valued quantum states of a qudit along the $n$ axes of the MAPS. Our future work will focus on utilizing the MAPS and these $d$-valued phase axial-based gates to visually build and cost-effectively construct useful $d$-valued quantum operators, such as arithmetic circuits, comparators, and counters, without requiring any $d^{\otimes m} \times d^{\otimes m}$ matrices multiplication, as we discussed such a visual-based technique using our ``Bloch sphere approach (BSA)'' \cite{albayaty2026bsa, albayaty2025bsa2} to construct useful cost-effective 2-valued quantum operators \cite{albayaty2026sxgrover, albayaty2025pswap, albayaty2025pswap2, albayaty2024toffoli, albayaty2024cala, albayaty2023gala, albayaty2025phdthesis}, where $d \geq 3$ and $m$ is the total number of qudits. Finally, as a generalized visualization $S^2$ framework, the MAPS could also be used for visualizing high-dimensional data for scientific and engineering applications, such as machine learning, quantum machine learning, quantum chemistry, just to name a few, where every spatial axis of the MAPS represents a single feature (dimensionality) of such data associated with its corresponding distinct values (numerical or textual).

\section{Background and Related Work}
To visualize higher $d$-valued quantum state-space of a qudit, where $d \geq 3$, physicists and mathematicians have introduced several visualization frameworks, including but not limited to the Bloch geometry, density operators and convex state-space, Majorana stellar constellations \cite{majorana1932, romero2024majorana, serrano2020majorana, bjork2015majorana}, and phase-space representations.

\subsection{Bloch geometry}
The formal geometrization of quantum state-space began with Felix Bloch's investigation of nuclear induction and magnetic resonance physics in 1946 \cite{bloch1946, bloch1946nuclear}, by introducing a three-dimensional vector space to describe the macroscopic spin dynamics of atomic nuclei. This was mathematically grounded in the works of Fano \cite{fano1957} and Landau \cite{landau1927}, who formalized the density operator $\rho$ to statistically describe mixed ensembles. For a 2-valued quantum system ($d = 2$), the $\rho$ is a positive semi-definite ($\rho \geq 0$), Hermitian operator \cite{nielsen2010, kaye2006, lapierre2021} acting on the two-dimensional Hilbert space ($\mathcal{H}_2 \cong \mathbb{C}^2$) \cite{young1988introduction, akhiezer2013theory}, and satisfying $\Tr(\rho) = 1$.

The Hilbert space $\mathcal{H}_d$, as a $d$-dimensional and complex-valued space representation ($d \geq 2$), plays an essential role in formalizing the $d$-valued quantum state-space. The contemporary quantum state-space visualization frameworks provide a bridge between the $\mathcal{H}_d$ formalism and the physically interpretable quantum physics and information science. A quantum state $|\psi\rangle$ in a $\mathcal{H}_d$ is described as stated in Eq.~(\ref{eq:psi}), where $\alpha_i$ is the quantum amplitude of the $i$th state, $\alpha_i \in \mathbb{C}$, $i \in \mathbb{Z}$, and $\sum_{i=0}^{d-1} |\alpha_i|^2=1$.

\begin{equation}
  |\psi\rangle=\sum_{i=0}^{d-1} \alpha_i |i\rangle
  \label{eq:psi}
\end{equation}

Notice that since the global phase $\gamma$ of $|\psi\rangle$ is physically irrelevant, i.e, $|\psi\rangle \approx e^{i\gamma}|\psi\rangle$, and all states belong to the complex projective geometry $\mathbb{CP}^{d-1} \cong S^{2d-1}/U(1)$. The modern geometrical frameworks for visualizing the quantum state-space have emerged from Bloch's spin ($\theta$ and $\phi$) formalism for a qubit ($d = 2$) \cite{bloch1946, bloch1946nuclear, bloch1953}, as expressed in Eq.~(\ref{eq:spin}).

\begin{equation}
  |\psi\rangle=\cos \left( \frac{\theta}{2} \right) |0\rangle + e^{i \phi} \sin \left( \frac{\theta}{2} \right) |1\rangle
  \label{eq:spin}
\end{equation}

Such a formalism is then mapped onto the unit $S^2$ sphere, i.e., the Bloch sphere, as stated in Eq.~(\ref{eq:rho}), where $\rho$ is the density operator, $\sigma$ denotes the Pauli vector of $(\sigma_x, \sigma_y, \sigma_z)^T$, $I_2$ is the $2 \times 2$ identity matrix, $r=(x,y,z)$ is the Bloch vector satisfying $||r|| \leq 1$, and $r \in \mathbb{R}^3$. The pure states corresponding to the boundary $\|r\| = 1$ form the unit $S^2$ sphere. While the mixed states occupy the interior $\|r\| < 1$, with the maximally mixed state $\rho = \frac{1}{2}I_2$ positioned precisely at the origin $r = 0$. Notice that the positive semi-definiteness condition $\rho \geq 0$ holds if and only if the eigenvalues $\lambda_1$ and $\lambda_2$ of $\rho$ are non-negative, which directly translates to the geometric constraint, as stated in Eq.~(\ref{eq:lambdas}).

\begin{equation}
  \rho = \frac{1}{2}\left(I_2 + r \cdot \sigma\right) = \frac{1}{2}
  \begin{pmatrix} 1 + z & x - iy \\ x + iy & 1 - z
  \end{pmatrix}
  \label{eq:rho}
\end{equation}

\begin{equation}
  \lambda_1 \lambda_2 = \frac{1}{4}(1 - \|r\|^2) \geq 0 \implies \|r\| \leq 1
  \label{eq:lambdas}
\end{equation}

In general, the Bloch sphere representation naturally connects to the Hopf fibrations \cite{lyons2003elementary, mosseri2001geometry, urbantke2003hopf}. The normalized qubit in the Hilbert space forms $S^3$, while its physically irrelevant global phase $\gamma$ gives $S^2$, as expressed in Eq.~(\ref{eq:S3-S2}).

\begin{equation}
  S^3/U(1) \cong \mathbb{CP}^1\cong S^2
  \label{eq:S3-S2}
\end{equation}

From Eq.~(\ref{eq:S3-S2}), such a topological reduction explains why the visualization of a qubit admits a unit $S^2$ sphere for $d = 2$, while the higher $d$-dimensional visualization frameworks generally do not. For qudits ($d \geq 3$), the generalized coherence-vector representations employ the generators $\lambda_i$ of $SU(d)$, as stated in Eq.~(\ref{eq:generators}), where $I_d$ is the $d \times d$ identity matrix and $r \in \mathbb{R}^{d^2-1}$.

\begin{equation}
  \rho = \frac{1}{d}I_d + \frac{1}{2} \sum_{i=1}^{d^2-1} r_i \lambda_i
  \label{eq:generators}
\end{equation}

In recent decades, the visualization of ququadit ($d=4$) and high-radix frameworks has integrated tools from differential geometry and entanglement theory. For a ququadit, the 15-dimensional quantum state-space contains complex subsets of mixed (separable and entangled) states. Salles \textit{et al.} \cite{salles2008} and Goyal \textit{et al.} \cite{goyal2016} expanded the use of generalized stereographic projections and mapping onto nested Hopf fibrations ($S^3 \rightarrow S^7 \rightarrow S^4$), providing a pathway to visualize bipartite entanglement dynamics by tracking the geometric distortions of the Bloch body projections. These mathematical formalisms provide the analytic framework required to interpret experimental tomographic data in modern high-radix quantum information systems.

\subsection{Density operators and convex state-space}
The interior Bloch geometry motivates convex state-space descriptions central to quantum tomography, because pure states form extremal points. For qutrits ($d=3$), Gell-Mann generators $\lambda_i$ \cite{gellmann1962} replace Pauli vectors $\sigma$, as expressed in Eq.~(\ref{eq:rho_3}), where $\xi \in \mathbb R^8$.

\begin{equation}
  \rho = \frac{1}{3} I_3 + \frac{1}{\sqrt{3}} \sum_{i=1}^{8} \xi_i \lambda_i
  \label{eq:rho_3}
\end{equation}

Kimura \cite{kimura2003} and Byrd-Khaneja \cite{byrd2003} showed that physical qutrit states occupy only a constrained subset of $\mathbb R^8$, defined by positivity conditions involving Casimir invariants \cite{abellanas1975general, vsnobl2005class} and symmetric star products $3 ||\xi||^2 - 2\xi \cdot (\xi * \xi) \le1$. Notice that general qudit systems require $d^2-1$ parameters. The generalized Bloch body $\Omega_d$ formalism expands density matrices using $SU(d)$ generators $\Lambda_i$, as stated in Eq.~(\ref{eq:generators2}), where $\xi \in \mathbb R^{d^2-1}$.

\begin{equation}
  \rho=
  \frac{1}{d}I_d
  +
  \sqrt{\frac{d-1}{2d}}
  \sum_{i=1}^{d^2-1}
  \xi_i\Lambda_i
  \label{eq:generators2}
\end{equation}

Mathematically, positivity constraints create non-spherical convex bodies, where their complexity increases rapidly with increasing radix $d$. However, the projective geometry provides another perspective. Hence, physical pure states belong to $\mathbb{CP}^{d-1} = S^{2d-1}/U(1)$ naturally generalizes the Bloch geometry. Distances between states may be quantified through the Fubini--Study distance metric $D_{FS}$ \cite{cheng2010quantum, erdmenger2023complexity}, as stated in Eq.~(\ref{eq:DFS}).

\begin{equation}
  D_{FS}(\psi,\phi)= \arccos(|\langle\psi|\phi\rangle|)
  \label{eq:DFS}
\end{equation}

Notice that the methods of Hopf fibrations extended such a visualization toward multi-qubit systems. Two-qubit geometry involves higher-dimensional fibrations sensitive to entanglement structure. Such constructions partially preserve geometric intuition lost in tensor-product Hilbert spaces.

\subsection{Majorana stellar constellations}
Majorana introduced an alternative geometric representation for spin systems in 1932 \cite{majorana1932}. Symmetric spin-$J$ states correspond to $2J$ points on a sphere. The Majorana polynomial, as stated in Eq.~(\ref{eq:Majorana}), maps quantum amplitudes $\alpha$ into spherical stellar constellations on the unit $S^2$ sphere. In general, the Majorana stellar constellations represent a symmetric state of $N = d-1$ qudits, as a collection of $N$ points on the unit $S^2$ sphere, by mapping a high-dimensional state to a point-cloud configuration. Notice that the polynomial roots determine constellations over $S^2$, allowing spin-$J$ visualization. Modern works \cite{romero2024majorana, serrano2020majorana, bjork2015majorana} applied Majorana geometry to symmetric multi-qubit states and qutrit visualization.

\begin{equation}
  P(z)=\sum_{m=-J}^{J}
  (-1)^{J-m}
  \sqrt{\binom{2J}{J+m}}
  \alpha_m z^{J+m}
  \label{eq:Majorana}
\end{equation}

\subsection{Phase-space representations}
Phase-space representations constitute another major direction, by mapping discrete or continuous quantum states onto a periodic (or spherical) phase-space grid, where the quantum state features emerge as negative quasi-probabilities or interference ripples. Wigner functions $W$ \cite{wigner1932, wootters1987} represent density operators over discrete or continuous phase spaces, as expressed in Eq.~(\ref{eq:Wigner}). Notice that negative quasi-probabilities visually expose non-classicality, and discrete Wigner constructions generalize toward finite-dimensional Hilbert spaces and qudit systems.

\begin{equation}
  W(p,q)=
  \frac{1}{2\pi\hbar}
  \int
  e^{i q y / \hbar}
  \left \langle p - \frac y2\right|
  \rho
  \left| p + \frac y2\right\rangle
  dy.
  \label{eq:Wigner}
\end{equation}

Quantum information geometry introduced metric structures to effectively quantify distinguishability between two density operators $p$ and $q$, including the Hilbert--Schmidt distance $D_{HS}$ \cite{ozawa2000entanglement}, as stated in Eq.~(\ref{eq:DHS}), and Bures distance $D_B$ \cite{sommers2003bures}, as stated in Eq.~(\ref{eq:Bures}), where the fidelity satisfies $F(p,q)=
\left(
  \mathrm{Tr}\left[
    \sqrt{
      \sqrt{p}
      q
      \sqrt{p}
  }\right]
\right)^2$.

\begin{equation}
  D_{HS}(p,q)=
  \sqrt{\mathrm{Tr}[(p-q)^2]}
  \label{eq:DHS}
\end{equation}

\begin{equation}
  D_B(p,q)=
  \sqrt{
    2-2\sqrt{F(p,q)}
  }
  \label{eq:Bures}
\end{equation}

Recent efforts investigate qudit analogues of the Bloch sphere using Heisenberg-Weyl operators \cite{sharma2021}, multi-dimensional spherical decompositions, hybrid geometric-probabilistic embeddings, just to name a few. Visualization remains an active research area because scalable quantum technologies increasingly require interpretable representations for high-dimensional quantum information systems.

Future visualization frameworks will likely integrate various topological structures, convex geometries, information metrics, tensor-network abstractions, and machine-learning dimensionality reduction. The historical trajectory from Bloch geometry to generalized qudit manifolds illustrates that visualization is not merely pedagogical but fundamental for understanding high-radix quantum state-space structures.

\section{Methods}
\begin{figure}[H]
  \centering
  \includegraphics[width=0.57\textwidth]{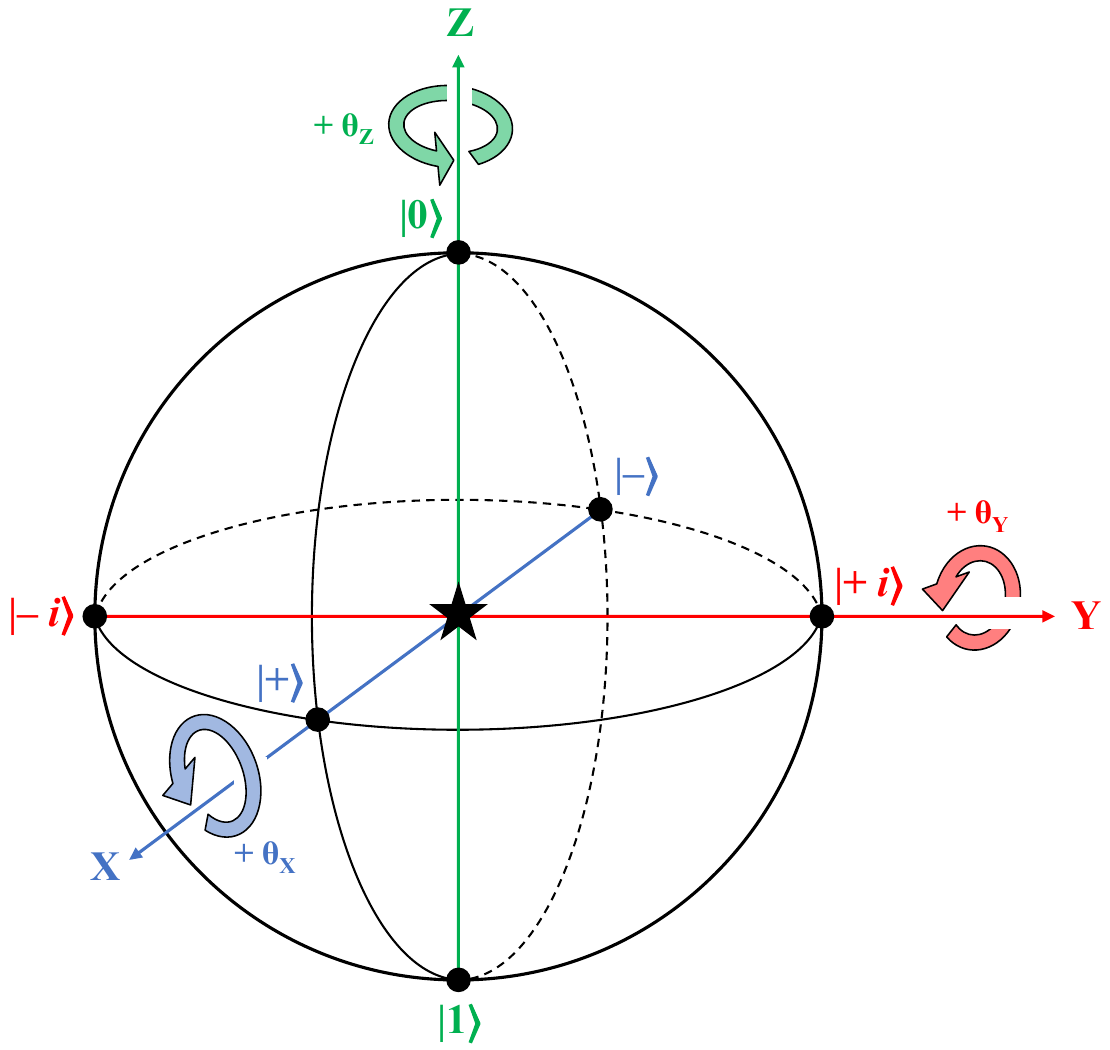}
  \caption{The Bloch sphere with three intersecting spatial axes (X, Y, and Z) for visualizing a qubit of: (i) six pure states (the circles) at the boundary $\|r\| = 1$, where the $|0\rangle$ and $|1\rangle$ states intersecting the Z-axis, the $|+\rangle$ and $|-\rangle$ states intersecting the X-axis, and the $|+i\rangle$ and $|-i\rangle$ states intersecting the Y-axis, and (ii) one maximally mixed state (the star) at the origin $r = 0$.}
  \label{fig:bloch}
\end{figure}

The Bloch sphere is a three-dimensional unit $S^2$ sphere with three intersecting spatial axes (X, Y, and Z) for visualizing the 2-valued pure and mixed states of a qubit ($d = 2$), as demonstrated in Fig.~\ref{fig:bloch}, where every spatial axis has its own rotational angle $\pm \theta$, the `$+\theta$' indicates a counterclockwise rotation, and the `$-\theta$' indicates a clockwise rotation.

As demonstrated in Fig.~\ref{fig:bloch}, the Z-axis represents the pure states of $|0\rangle$ and $|1\rangle$ in Dirac notation \cite{nielsen2010, kaye2006, lapierre2021}, which are equivalent to $\left[ 1,0 \right]^T$ or $\left[0,1\right]^T$ in Heisenberg notation \cite{nielsen2010, kaye2006, lapierre2021}, respectively. While the parallel planar projections \cite{albayaty2026bsa, albayaty2025pswap}, such as the equator of the Bloch sphere, intersected with the X-axis and Y-axis represent the mixed states of $\frac{1}{\sqrt{2}} \left(|0\rangle \pm j|1\rangle \right) = \frac{1}{\sqrt{2}} \left[1, \pm j1 \right]^T$, where $\pm j = e^{i \pi k} = \cos(\pi k) + i \sin (\pi k)$ and $k \in \mathbb{R}$.

Please observe that, in the Bloch sphere, (i) the global and relative (local) phases of the pure and mixed states are not directly visualized along these three spatial axes, and (ii) the pure $|0\rangle$ state is physically orthogonal to the pure $|1\rangle$ state separated by $\pi/2$ radians; however, these two states are visualized as non-orthogonal (normal) pure states on the Z-axis separated by $\pi$ radians! For these two contradictions, additional mathematical formalisms are required to accurately describe the orthonormality with the global and relative phases of the complete 2-valued quantum state-space of a qubit.

\subsection{Multi-axial projective sphere}
\begin{figure}[H]
  \centering
  \begin{subfigure}[b]{0.75\textwidth}
    \centering
    \includegraphics[width=0.75\textwidth]{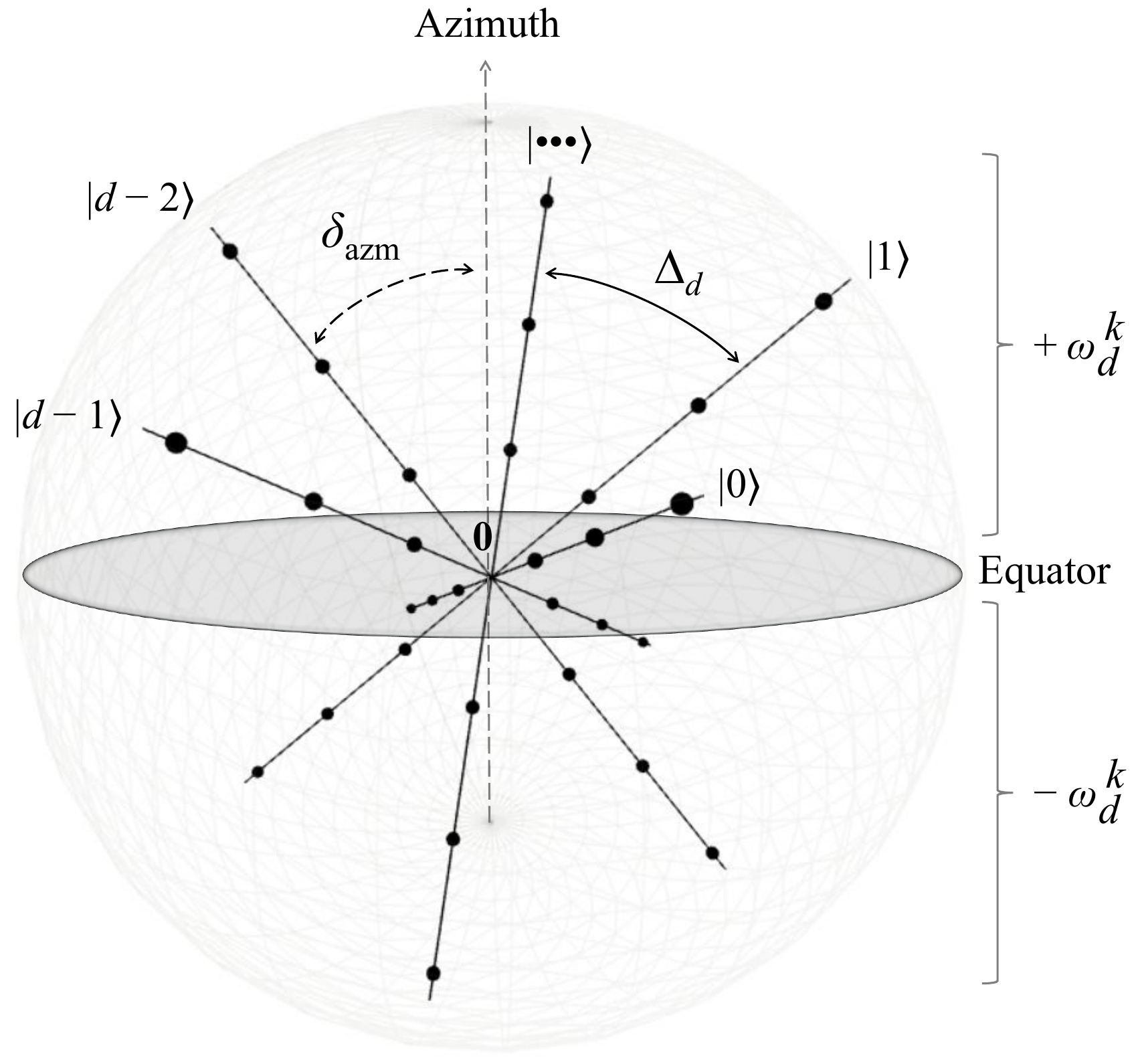}
    \caption{The MAPS as an $S^2$ framework}
  \end{subfigure}
  \vfill
  \begin{subfigure}[b]{0.9\textwidth}
    \centering
    \includegraphics[width=0.75\textwidth]{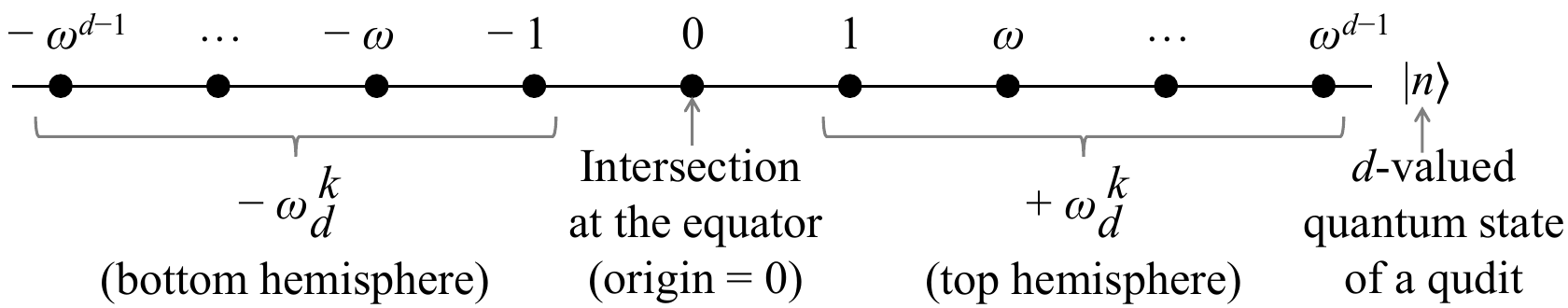}
    \caption{One spatial axis of the MAPS with a set of relative phases}
  \end{subfigure}
  \hfill
  \caption{The geometrical construction of our generalized MAPS for visualizing $d$-valued quantum states of a qudit using: \textbf{(a)} an $S^2$ framework,  and \textbf{(b)} $n$ intersecting spatial axes, where $d \geq 3$, $0 \leq n \leq d-1$, $\Delta_d = 2 \pi/d$ radians, $\delta_{\text{azm}} = \pi/z$ radians, $k \in \mathbb{Z}$, $z \in \mathbb{R}$, and the black circles indicate the distinct relative phases of $+\omega_d^k$ and $-\omega_d^k$.}
  \label{fig:generalized_MAPS}
\end{figure}

The work of Bloch \cite{bloch1946, bloch1946nuclear, bloch1953} provided a visualization framework with ease of illustration, structural simplicity, and natural representation of 2-valued quantum states of a qubit. Such a straightforward, simple visualization approach of Bloch influences us to similarly design a new generalized three-dimensional $S^2$ sphere for engineers and researchers to directly visualize higher $d$-valued quantum state-space of a qudit, including the notation of global and relative phases, where $d \geq 3$. We called this new generalized $S^2$ sphere the ``multi-axial projective sphere (MAPS)'', as presented in Definition 1 and shown in Fig.~\ref{fig:generalized_MAPS}.

\noindent
\textbf{Definition 1:} The multi-axial projective sphere (MAPS), as a generalized $d$-valued state-space visualization $S^2$ framework, is geometrically constructed with the following constraints and properties, where $d \geq 3$, $0 \leq n \leq d-1$, $k \in \mathbb{Z}$, and $z \in \mathbb{R}$.

\begin{enumerate}
  \item The $S^2$ framework consists of $n$ projectional intersecting spatial axes.

  \item Every spatial axis is mapped to one $d$-valued state-space, e.g., a $d$-valued quantum state of a qudit, which is labeled the ``$|n\rangle$''-axis on the $S^2$ framework.

  \item Every spatial axis has a set of distinct values $\pm \omega_d^k$, e.g., a set of relative (local) phases for that $d$-valued quantum state of a qudit, as stated in Eq.~(\ref{eq:omega-k}).

  \item All $n$ spatial axes are: (i) projectionally intersected at the origin = 0 for all $\pm \omega_d^k$, (ii) equally separated by the ``pseudo-orthogonal angle $\Delta_d$'' expressed in Eq.~(\ref{eq:Delta-d}), and (iii) uniformly displaced from the azimuth pole by the ``azimuth displacement angle $\delta_{\text{azm}}$'' stated in Eq.~(\ref{eq:delta-azm}).

  \item The ``equator'' divides the $S^2$ framework into: (i) a ``top hemisphere'' consisting of all $+\omega_d^k$ and (ii) a ``bottom hemisphere'' consisting of all $-\omega_d^k$, for all $n$ spatial axes.
\end{enumerate}

\begin{equation}
  \pm \omega_d^k = e^{i \frac{2 \pi k}{d}} = \cos(\frac{2 \pi k}{d}) + i \sin(\frac{2 \pi k}{d})
  \label{eq:omega-k}
\end{equation}

\begin{equation}
  \Delta_d = \frac{2 \pi}{d}
  \label{eq:Delta-d}
\end{equation}

\begin{equation}
  \delta_{\text{azm}} = \frac{\pi}{z}
  \label{eq:delta-azm}
\end{equation}

\noindent
$\square$ \\

Please observe that, in Definition 1, the orthogonality across all $|n\rangle$ axes of the MAPS is not defined by the conventional perpendicular angle ``$\pi/2$'' radians. However, in this research, we propose the ``pseudo-orthogonal angle $\Delta_d$'' for every $|n\rangle$-axis, as stated in Eq.~(\ref{eq:Delta-d}) above, and the ``orthogonality completeness $\Delta_\text{completeness}$'' across all $|n\rangle$ axes is fulfilled when all $\Delta_d$ rotationally occupy the entire boundary of the $S^2$ framework, i.e., $\Delta_\text{completeness}$ reaches ``$2\pi$'' radians, as expressed in Eq.~(\ref{eq:orthogonality-completeness}). Notice that when $\Delta_\text{completeness}$ is not fulfilled, all $|n\rangle$ axes of the MAPS are not separated evenly by $\Delta_d$, yielding to visually indistinguishable and geometrically overlapped axes.

\begin{equation}
  \Delta_\text{completeness} = \sum_{d} \Delta_d = \sum_{d} \frac{2\pi}{d} = 2\pi~\text{radians}
  \label{eq:orthogonality-completeness}
\end{equation}

\begin{figure}[H]
  \centering
  \begin{subfigure}[b]{0.49\textwidth}
    \centering
    \includegraphics[width=\textwidth]{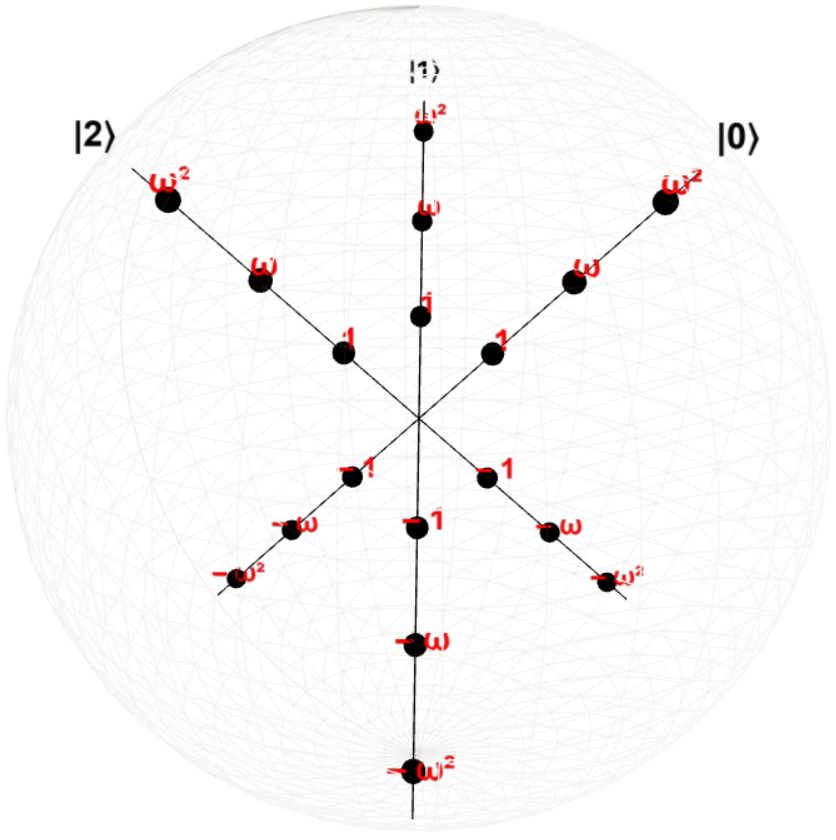}
    \caption{MAPS for a qutrit ($d = 3$)}
  \end{subfigure}
  \vfill
  \begin{subfigure}[b]{0.49\textwidth}
    \centering
    \includegraphics[width=\textwidth]{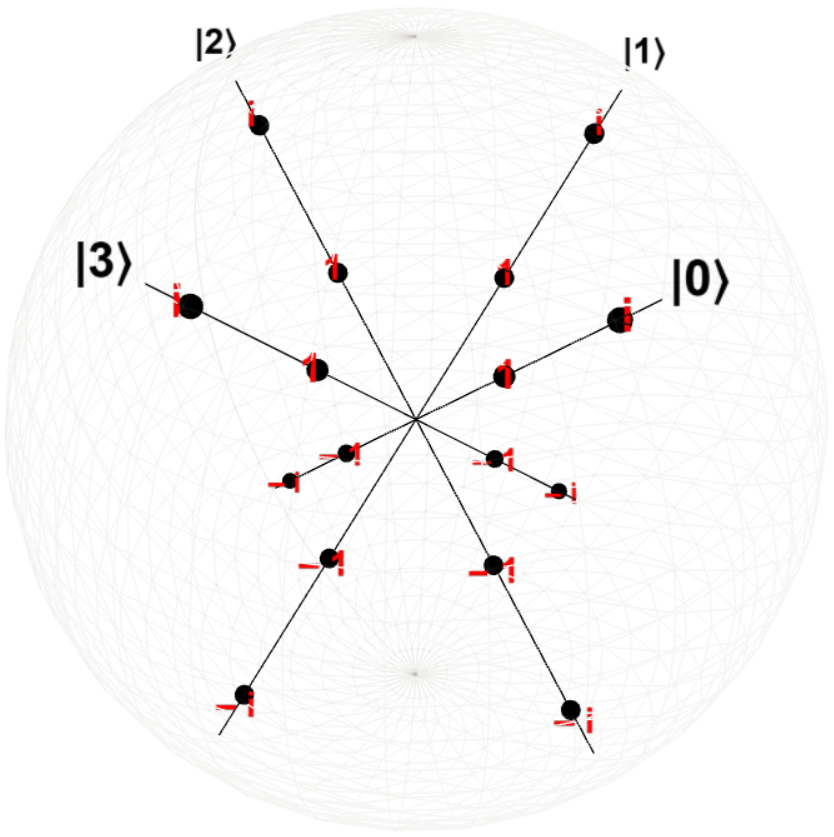}
    \caption{MAPS for a ququadit ($d = 4$)}
  \end{subfigure}
  \hfill
  \begin{subfigure}[b]{0.49\textwidth}
    \centering
    \includegraphics[width=\textwidth]{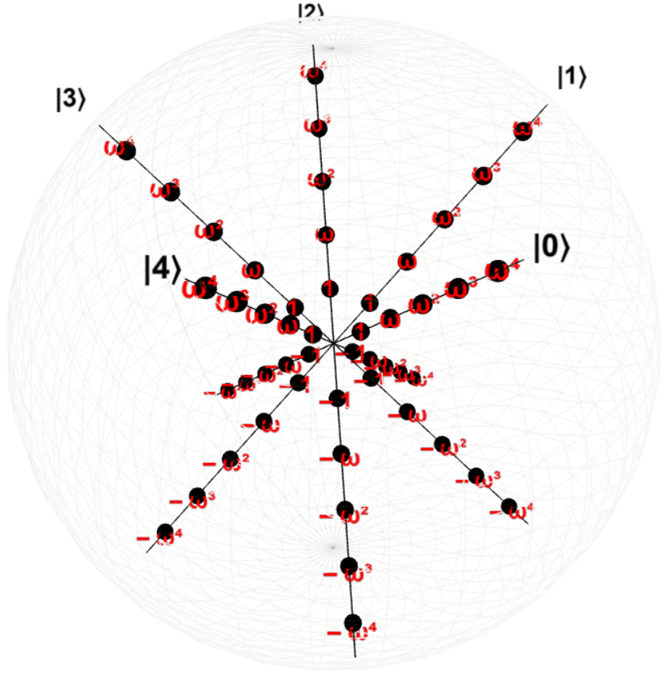}
    \caption{MAPS for a quintit ($d = 5$)}
  \end{subfigure}
  \caption{Three MAPS visualizing: \textbf{(a)} the 3-valued quantum states of a qutrit, where $\pm \omega_3^k = \{\pm 1, \pm \omega, \pm \omega^2\}$ and $\Delta_3 = 2\pi/3$ radians, \textbf{(b)} the 4-valued quantum states of a ququadit, where $\pm \omega_4^k = \{\pm 1, \pm i\}$ and $\Delta_4 = \pi/2$ radians, and \textbf{(c)} the 5-valued quantum states of a quintit, where $\pm \omega_5^k = \{\pm 1, \pm \omega, \pm \omega^2, \pm \omega^3, \pm \omega^4\}$ and $\Delta_5 = 2\pi/5$ radians. Notice that $\delta_{\text{azm}} = \pi/4$ radians for all three MAPS \cite{albayaty2026maps}.}
  \label{fig:examples_MAPS}
\end{figure}

For instance, for the 3-valued quantum states of a qutrit ($d = 3$), the MAPS consists of three projectional intersecting spatial axes labeled $|0\rangle$, $|1\rangle$, and $|2\rangle$, equally separated by $\Delta_3 = 2\pi/3$ radians, and uniformly displaced by $\delta_{\text{azm}} = \pi/4$ radians, with every spatial axis having the set of relative phases $\pm \omega_3^k = \{\pm 1, \pm  \omega, \pm  \omega^2\}$, as shown in Fig.~\ref{fig:examples_MAPS}(a). Notice that, in Fig.~\ref{fig:examples_MAPS}(a), $\Delta_\text{completeness} = \sum_{3} \Delta_3 = 2\pi$ radians, i.e., the orthogonality is fulfilled.

Moreover, Fig.~\ref{fig:examples_MAPS}(b) and Fig.~\ref{fig:examples_MAPS}(c) demonstrate two MAPS visualizing the 4-valued quantum states of a ququadit ($d = 4$) and the 5-valued quantum states of a quintit ($d = 5$), respectively. The open-source 3D viewer project of MAPS is available at \cite{albayaty2026maps}, and this 3D viewer supports the orbital geometries and visualization controls of rotation, zooming in/out, and panning.

In our research, we noticed that the (pure or mixed) $|0\rangle$ state of a qudit ($d \geq 3$) always has a relative phase of 1. For this reason, the relative phase for the $|0\rangle$-axis of the MAPS always indicates the global phase for the overall $d$-valued quantum states of a qudit. Therefore, the MAPS has an advantageous visualization feature compared to the Bloch sphere, by directly visualizing the global and relative phases of the $d$-valued quantum states along all $n$ spatial axes, without requiring any mathematical formalism to describe the complete $d$-valued quantum state-space of a qudit.

\begin{figure}[H]
  \centering
  \begin{subfigure}[b]{0.49\textwidth}
    \centering
    \includegraphics[width=\textwidth]{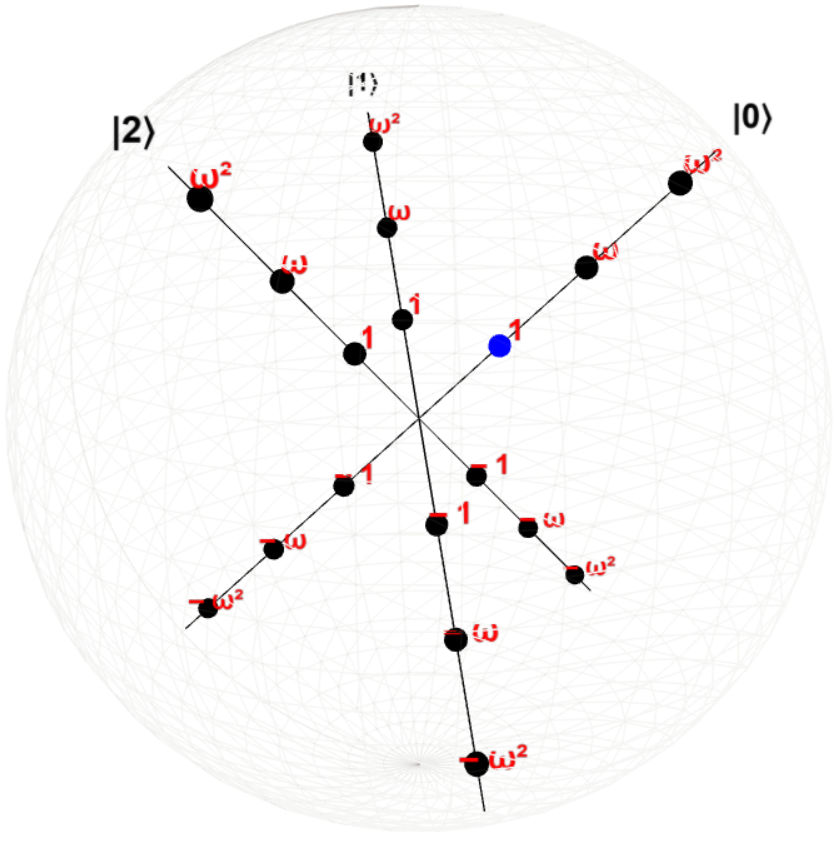}
    \caption{Pure $|0\rangle$ state with a global phase = 1}
  \end{subfigure}
  \hfill
  \begin{subfigure}[b]{0.49\textwidth}
    \centering
    \includegraphics[width=\textwidth]{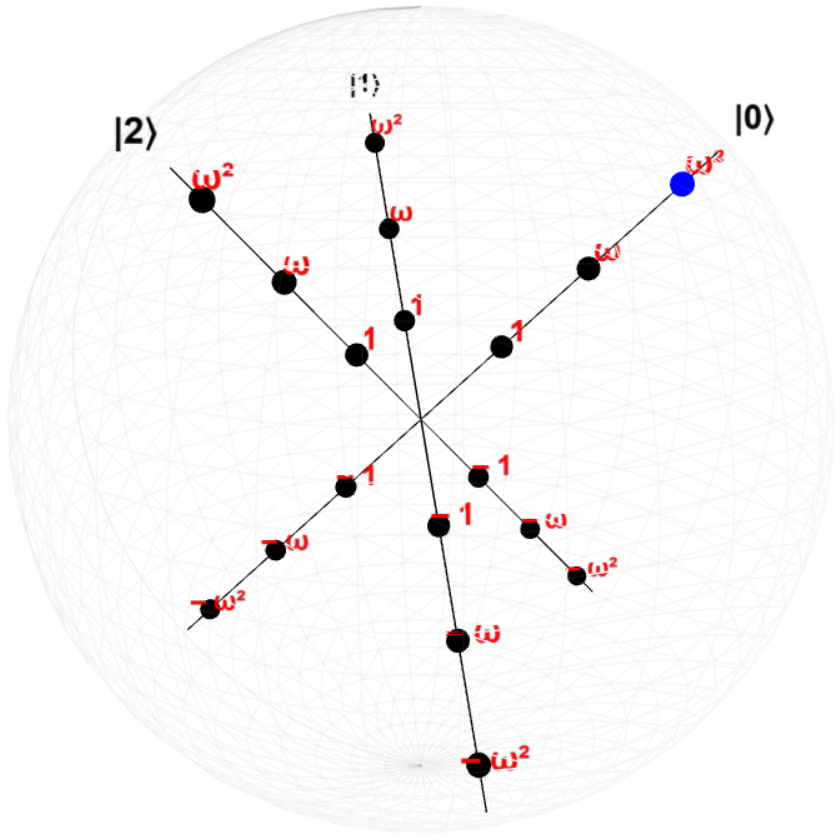}
    \caption{Pure $|0\rangle$ state with a global phase = $\omega^2$}
  \end{subfigure}
  \vfill
  \begin{subfigure}[b]{0.49\textwidth}
    \centering
    \includegraphics[width=\textwidth]{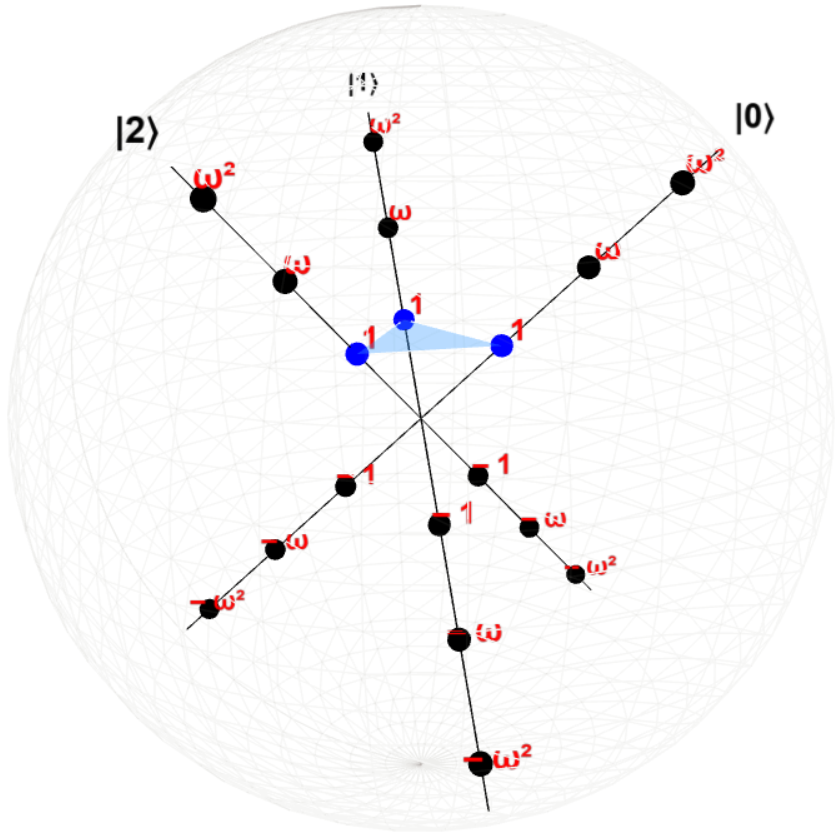}
    \caption{Mixed states with a global phase = 1}
  \end{subfigure}
  \hfill
  \begin{subfigure}[b]{0.49\textwidth}
    \centering
    \includegraphics[width=\textwidth]{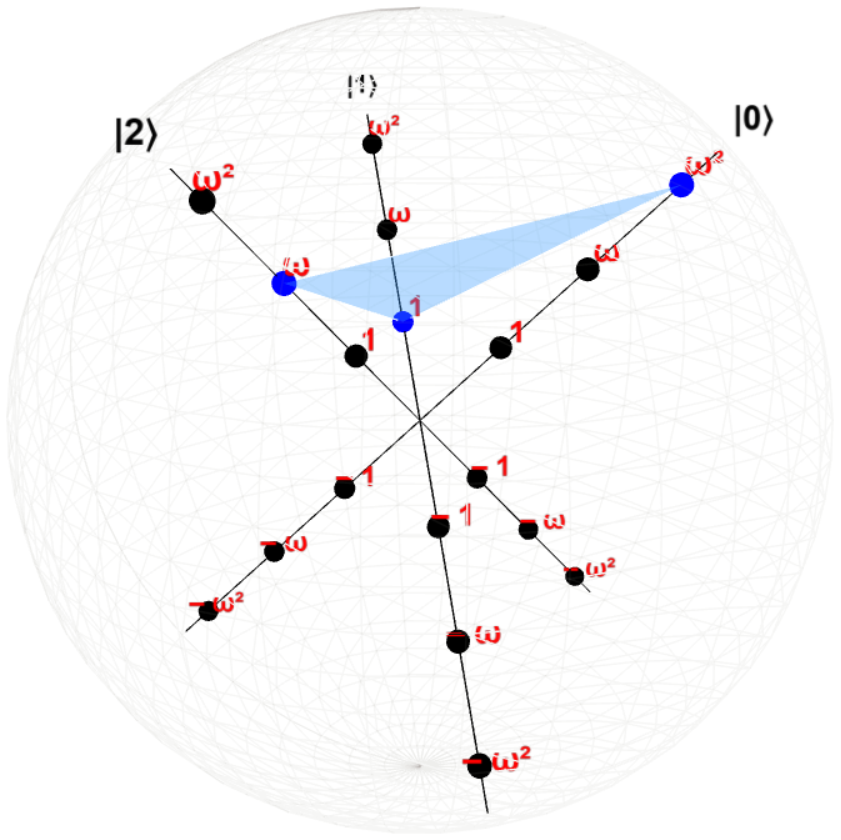}
    \caption{Mixed states with a global phase = $\omega^2$}
  \end{subfigure}
  \caption{Four MAPS visualizing the pure and mixed states of qutrits ($d = 3$) with the global phases of: \textbf{(a)} 1, i.e., $|0\rangle = [1,0,0]^T$, \textbf{(b)} $\omega^2$, i.e., $\omega^2|0\rangle = [\omega^2,0,0]^T = \omega^2[1,0,0]^T$, \textbf{(c)} 1, i.e., $\frac{1}{\sqrt{3}} \left(|0\rangle + |1\rangle + |2\rangle \right) = \frac{1}{\sqrt{3}} [1,1,1]^T$, and \textbf{(d)} $\omega^2$, i.e., $\frac{1}{\sqrt{3}} \left( \omega^2|0\rangle + |1\rangle + \omega|2\rangle \right) = \frac{1}{\sqrt{3}} [\omega^21,1,\omega1]^T = \frac{\omega^2}{\sqrt{3}} [1,\omega1,\omega^21]^T$, where the blue circles indicate global and relative phases, and the translucent blue triangles indicate the 3-valued mixed state-space \cite{albayaty2026maps}.}
  \label{fig:global_phases_0_state}
\end{figure}

For instance, for qutrits ($d=3$), Fig.~\ref{fig:global_phases_0_state}(a) shows the pure $|0\rangle$ state with the global phase of 1, which is equivalent to $|0\rangle = [1,0,0]^T$, Fig.~\ref{fig:global_phases_0_state}(b) illustrates the pure $|0\rangle$ state with the global phase of $\omega^2$, which is equivalent to $\omega^2|0\rangle = [\omega^2,0,0]^T = \omega^2[1,0,0]^T$, Fig.~\ref{fig:global_phases_0_state}(c) demonstrates the mixed states with the global phase of 1 (without any relative phase), which is equivalent to $\frac{1}{\sqrt{3}}\left( |0\rangle + |1\rangle + |2\rangle \right) = \frac{1}{\sqrt{3}} [1,1,1]^T$, and Fig.~\ref{fig:global_phases_0_state}(d) depicts the mixed states with the global phase of $\omega^2$ (with relative phases for $|1\rangle$ and $|2\rangle$), which is equivalent to $\frac{1}{\sqrt{3}}\left( \omega^2|0\rangle + |1\rangle + \omega|2\rangle \right) = \frac{1}{\sqrt{3}} [\omega^21,1,\omega1]^T = \frac{\omega^2}{\sqrt{3}} [1,\omega1,\omega^21]^T$.

\begin{figure}[H]
  \centering
  \includegraphics[width=0.6\textwidth]{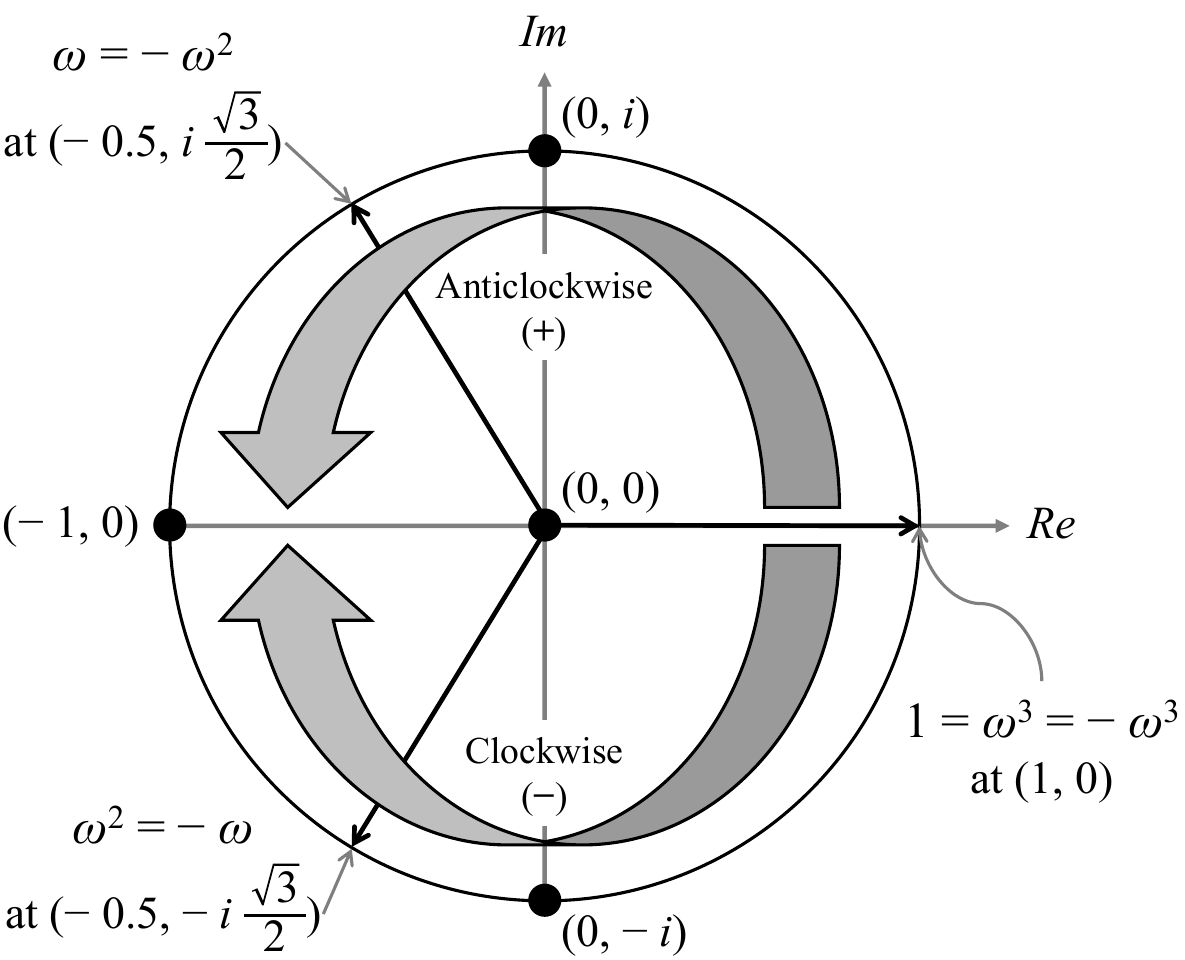}
  \caption{The two-dimensional phase plane (a real-imaginary unit circle) to calculate the phase periodicity for a qutrit ($d = 3$) of $\pm \omega_3^k = \omega_3^{3+\alpha} = \omega_3^\alpha = -\omega_3^{3-\beta}$, where $k = 3 + \alpha$, $\beta = k~\text{mod}~3$, and $\forall k, \alpha, \beta \in \mathbb{Z}$.}
  \label{fig:phase_plane}
\end{figure}

Our geometrical representation of the MAPS is considered a straightforward technique for implicitly visualizing the global phases, relative phases, and axial rotation operations applied to a qudit, as compared to the Bloch sphere ($d=2$) and the indistinct two-dimensional phase plane ($d=3$), which is a real-imaginary unit circle, as depicted in Fig.~\ref{fig:phase_plane}. Notice that, in Fig.~\ref{fig:phase_plane} for $d=3$, by using the periodic geometry technique, $1 + \omega_3 + \omega_3^2 = 0$, $\omega_3^3 = 1 = -\omega_3^3$, $\omega_3^4 = \omega_3 = -\omega_3^2$, $\omega_3^5 = \omega_3^2 = -\omega_3$, and so on. Hence, the phase periodicity of $\pm \omega_3^k$ is expressed in Eq.~(\ref{eq:phase-periodicity}), where $k = 3 + \alpha$ and $\beta = k~\text{mod}~3$.

\begin{equation}
  \pm \omega_3^k = \omega_3^{3+\alpha} = \omega_3^\alpha = -\omega_3^{3-\beta}; \forall k, \alpha, \beta \in \mathbb{Z}
  \label{eq:phase-periodicity}
\end{equation}

Therefore, for a qudit ($d \geq 3$), the phase periodicity of $\pm \omega_d^k$ is rotationaly generalized in Eq.~(\ref{eq:phase-periodicity-d}), where $k = d + \alpha$ and $\beta = k~\text{mod}~d$.

\begin{equation}
  \pm \omega_d^k = \omega_d^{d+\alpha} = \omega_d^\alpha = -\omega_d^{d-\beta}; \forall k, \alpha, \beta \in \mathbb{Z}
  \label{eq:phase-periodicity-d}
\end{equation}

Please observe that the two-dimensional phase plane, as an $S^1$ framework, is: (i) only applicable when $d=3$, but not applicable when $d \geq 4$ due to the angular orthogonality inconsistency, and (ii) not physically representing the correct phase periodicity rotations. For instance, in Fig.~\ref{fig:phase_plane}, the phases $\omega_3$ and $-\omega_3^2$ are not physically equivalent in rotations; however, they are geometrically correct, as expressed in Eq.~(\ref{eq:phase-periodicity}) above! Hence, there is a visualization contradiction between the physical interpretation and the geometrical representation for describing the complete 3-valued quantum state-space of a qutrit.

Subsequently, our MAPS resolves such orthogonality inconsistency and visualization contradiction by proposing the $\Delta_d$ and $\Delta_\text{completeness}$ formalisms through providing a unique projectional axis for every $d$-valued quantum state of a qudit ($d \geq 3$), with separable sets of $\pm \omega_d^k$ by the equator of the MAPS, for visually clarifying the physical operations to match their corresponding geometrical rotations, as presented in Definition 1 above, stated in Eq.~(\ref{eq:phase-periodicity-d}) above, illustrated in Fig.~\ref{fig:generalized_MAPS}, and discussed next using our proposed group of novel $d$-valued phase-axial swiveling and shifting gates $PASS_d$. Therefore, the MAPS implicitly preserves the quantum properties, visual distinguishability, entanglement structures, and unitary dynamics for all $d$-valued quantum (pure and mixed) states of a qudit.

\subsection{Novel phase-axial swiveling and shifting gates}
For $d \geq 3$, the conventional well-known $d$-valued quantum gates have been thoroughly discussed in the literature \cite{moraga2014some, cambou2018can, khan2005synthesis, wang2011improved, yang2005realizing, albayaty2026ternary}, such as the ternary Chrestenson ($CH$) superposition gate, inverse $CH$ gate ($CH^\dagger$), permutative gates ($01$, $02$, and $12$), shift gates ($+1$ and $+2$), $d$-valued phase gate ($Z_d$), inverse $Z_d$ gate ($Z_d^\dagger$), just to name a few. Based on our introduced MAPS visualization framework, a group of novel $d$-valued phase axial-based gates is proposed in this paper. We called this phase-based gate the ``$d$-valued phase-axial swiveling and shifting gate $PASS_d$'', as presented in Definition 2. In general, this group consists of $|\pm \omega_d^k |^d-1$ permutative $PASS_d$ gates, where $| \pm \omega_d^k |$ is the number of all relative phases ($+\omega_d^k$ and $-\omega_d^k$), i.e., the cardinality of $\pm \omega_d^k$. \\

\noindent
\textbf{Definition 2:} The $d$-valued phase-axial swiveling and shifting gate $PASS_d$ is a $d \times d$ diagonal unitary operator that swivels (rotates) and shifts (scales) the global and relative (local) phases of $d$-valued (pure and mixed) states for a qudit along the $|n\rangle$ axes in both hemispheres of the MAPS, as expressed in Eq.~(\ref{eq:PASS_d}), where $d \geq 3$, $\forall \theta_{|n\rangle} \in \pm \omega_d^k$, $\pm \omega_d^k = e^{i \frac{2 \pi k}{d}} = \cos(\frac{2 \pi k}{d}) + i \sin(\frac{2 \pi k}{d})$, $k \in \mathbb{Z}$, and $0 \leq n \leq d-1$. For the overall $d$-valued quantum state-space of a qudit, the global and relative phases are swiveled, i.e., axially rotated from one hemisphere to another hemisphere, and shifted, i.e., axially scaled in the same hemisphere, with the following $\theta$ angular constraints.

\begin{enumerate}
  \item $\forall \theta_{|n\rangle} \in \{1\} \rightarrow$ $PASS_d = I_d$, which is the $d \times d$ identity matrix, i.e., nothing occurs.

  \item $\forall \theta_{|n \neq 0\rangle} \in \{1\}$ and $\theta_{|0\rangle} \in -\omega_d^k \rightarrow$ The global phase is swiveled.

  \item $\forall \theta_{|n \neq 0\rangle} \in \{1\}$ and $\theta_{|0\rangle} \in +\omega_d^k \setminus \{1\} \rightarrow$ The global phase is shifted.

  \item $\exists \theta_{|n \neq 0\rangle} \in -\omega_d^k$ and $\theta_{|0\rangle} \in \{1\} \rightarrow$ A set of relative phases is swiveled.

  \item $\exists \theta_{|n \neq 0\rangle} \in +\omega_d^k \setminus \{1\}$ and $\theta_{|0\rangle} \in \{1\} \rightarrow$ A set of relative phases is shifted.

  \item $\exists \theta_{|n\rangle} \in -\omega_d^k \rightarrow$ The global phase and/or a set of relative phases are swiveled.

  \item $\exists \theta_{|n\rangle} \in +\omega_d^k \setminus \{1\} \rightarrow$ The global phase and/or a set of relative phases are shifted.

  \item $\forall \theta_{|n\rangle} \in \pm \omega_d^k \setminus \{1\} \rightarrow$ All global and relative phases are both swiveled and shifted.
\end{enumerate}

\begin{equation}
  PASS_d =
  \left[
    \begin{array}{llll}
      \theta_{|0\rangle} & 0 & \cdots & 0 \\
      0 & \theta_{|1\rangle} & \cdots & 0 \\
      \vdots & \vdots & \ddots & \vdots \\
      0 & 0 & \cdots & \theta_{|d-1\rangle}
    \end{array}
  \right]
  \label{eq:PASS_d}
\end{equation}

\noindent
$\square$ \\ \\

Please observe the three essential quantum unitary dynamics of the $PASS_d$ gate as follows.

\begin{enumerate}
  \item The $PASS_d$ is a unitary and Hermitian \cite{nielsen2010, kaye2006, lapierre2021} gate, iff $\forall \theta_{|n\rangle} \in \{-1,1\}$. Otherwise, the $PASS_d$ is a unitary and non-Hermitian gate, and its $PASS_d^\dagger$ gate can be directly constructed by inversing every $\theta_{|n\rangle}$ through: (i) inverting only the signs of all $i$'s when $d = 4$ and (ii) reversing only the phase periodicity of all $\omega$'s (with identical signs) when $d \neq 4$, as stated in Eq.~(\ref{eq:phase-periodicity}) and Eq.~(\ref{eq:phase-periodicity-d}) above. Such that, in constructing a $PASS_d^\dagger$ gate for a $PASS_d$ gate, $i \Leftrightarrow -i$ and $\pm \omega_d^k \Leftrightarrow \pm \omega_d^\alpha$, where $d = k - \alpha$ and $\forall k, \alpha, \beta \in \mathbb{Z}$.

  \item For a group of $|\pm \omega_d^k |^d-1$ permutative $PASS_d$ gates: (i) when $d = 3$, this group has 215 permutative $PASS_3$ and $PASS_3^\dagger$ gates, excluding $I_3$, (ii) when $d = 4$, this group has 255 permutative $PASS_4$ and $PASS_4^\dagger$ gates, excluding $I_4$, and so on.

  \item All $PASS_d$ and $PASS_d^\dagger$ gates are considered the generalized phase-based gates of the well-known $Z_d$ gate \cite{moraga2014some, albayaty2026ternary, pudda2024generalised, goss2022high} expressed in Eq.~(\ref{eq:z_d}), since any $\theta_{|n\rangle}$ in a $PASS_d$ or $PASS_d^\dagger$ gate can be individually customized for desirable swiveling and/or shifting quantum rotational operations.

    \begin{equation}
      Z_d =
      \left[
        \begin{array}{llll}
          1 & 0 & \cdots & 0 \\
          0 & \omega & \cdots & 0 \\
          \vdots & \vdots & \ddots & \vdots \\
          0 & 0 & \cdots & \omega^{d-1}
        \end{array}
      \right]
      \label{eq:z_d}
    \end{equation}
\end{enumerate}

\section{Results}
The potential competence of MAPS is extremely evident when visualizing $d$-valued mixed (superimposed) states of a qudit, using $n$ projective spatial axes for every $|n\rangle$ pure state associated with a set of $PASS_d$ gates, where $d \geq 3$ and $0 \leq n \leq d-1$. For ease of illustration and brevity, $d = 3$ is mostly used here to geometrically visualize the 3-valued mixed states of a qutrit, as discussed and demonstrated in the following examples. \\

\begin{figure}[H]
  \centering
  \begin{subfigure}[b]{0.49\textwidth}
    \centering
    \includegraphics[width=\textwidth]{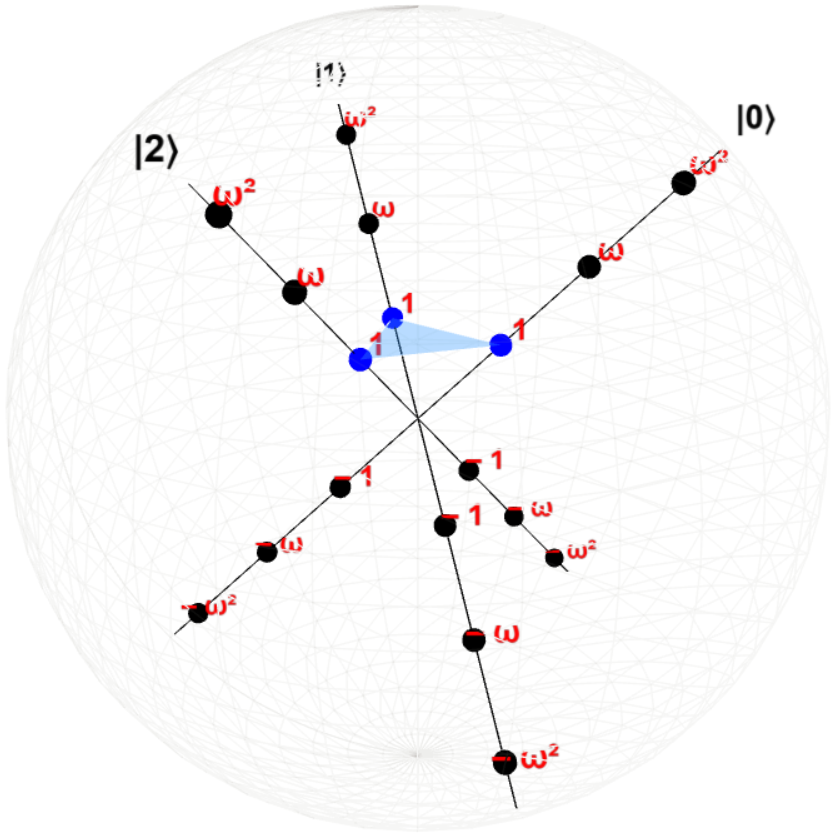}
    \caption{ $\frac{1}{\sqrt{3}}\left( |0\rangle + |1\rangle + |2\rangle \right) = \frac{1}{\sqrt{3}}[1,1,1]^T $ }
  \end{subfigure}
  \hfill
  \begin{subfigure}[b]{0.49\textwidth}
    \centering
    \includegraphics[width=\textwidth]{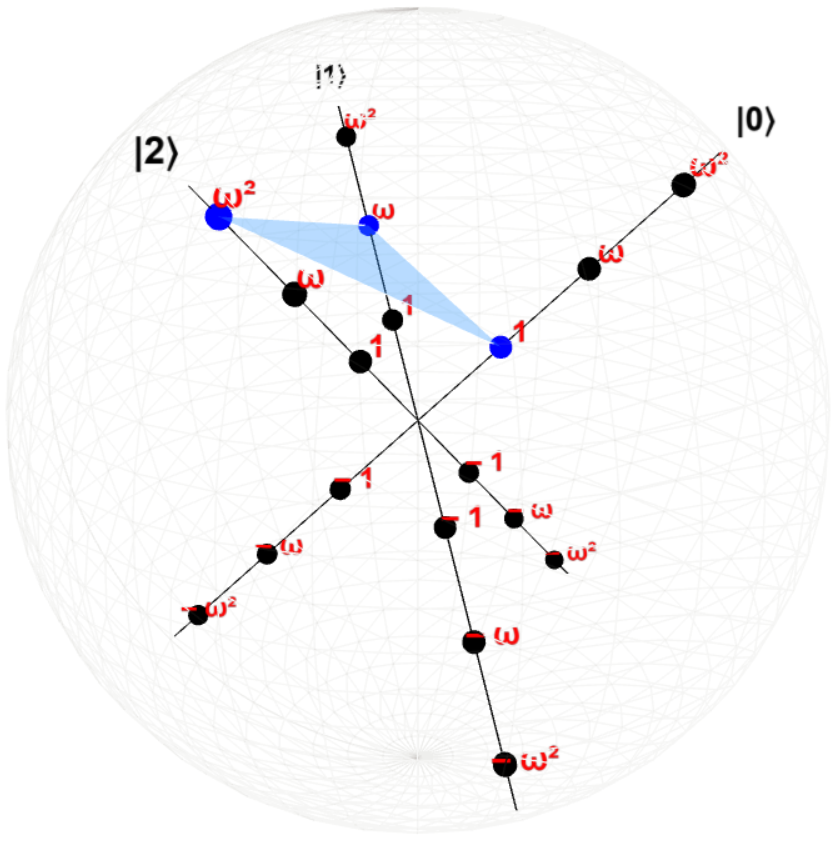}
    \caption{ $\frac{1}{\sqrt{3}}\left( |0\rangle + \omega|1\rangle + \omega^2|2\rangle \right) = \frac{1}{\sqrt{3}}[1,\omega,\omega^2]^T $ }
  \end{subfigure}
  \vfill
  \begin{subfigure}[b]{0.49\textwidth}
    \centering
    \includegraphics[width=\textwidth]{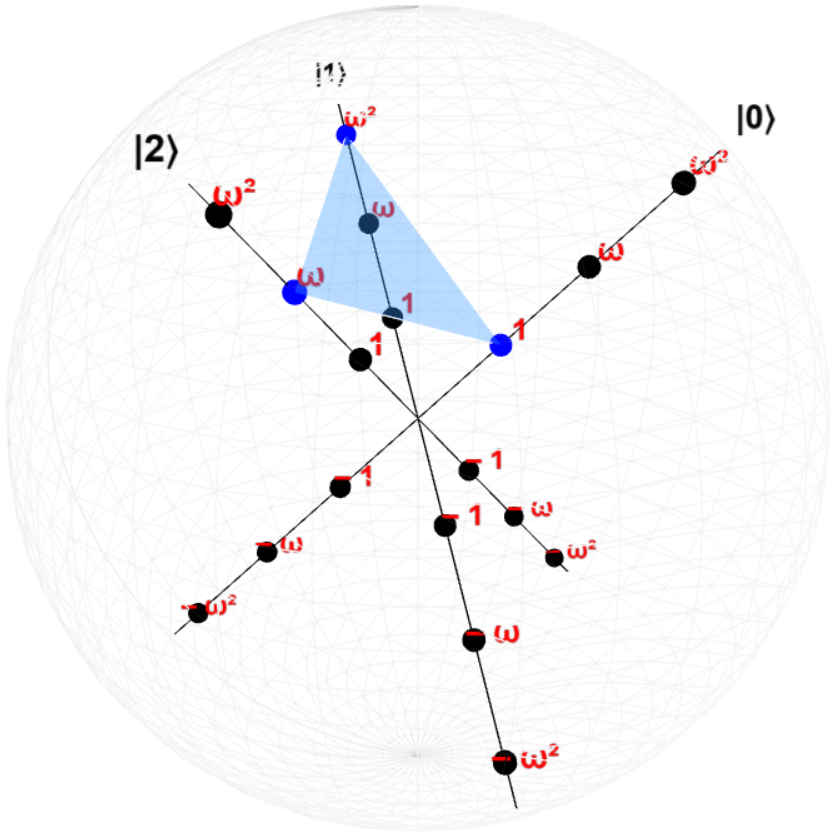}
    \caption{ $\frac{1}{\sqrt{3}}\left( |0\rangle + \omega^2|1\rangle + \omega|2\rangle \right) = \frac{1}{\sqrt{3}}[1,\omega^2,\omega]^T $ }
  \end{subfigure}
  \caption{Three MAPS visualizing the mixed states of three qutrits after applying the $CH$ gate to the pure states of: \textbf{(a)} $|0\rangle$, \textbf{(b)} $|1\rangle$, and \textbf{(c)} $|2\rangle$, where the blue circles indicate relative phases and the translucent blue triangles indicate the 3-valued mixed states \cite{albayaty2026maps}.}
  \label{fig:case_1}
\end{figure}

\noindent
\textbf{Example 1: Superposition gates} \\
For $d = 3$, the well-known ternary Chrestenson ($CH$) superposition gate \cite{moraga2014some, albayaty2026ternary, chrestenson1955class, mandal2014synthesis, al2002multiple}, as stated in Eq.~(\ref{eq:chrestenson}), is applied to three qutrits of initial pure states of $|0\rangle$, $|1\rangle$, and $|2\rangle$ to create three mixed states, as shown in Fig.~\ref{fig:case_1}(a), Fig.~\ref{fig:case_1}(b), and Fig.~\ref{fig:case_1}(c), respectively.

\begin{equation}
  CH =
  \frac{1}{\sqrt{3}} \left[
    \begin{array}{lll}
      1 & 1 & 1 \\
      1 & \omega & \omega^2 \\
      1 & \omega^2 & \omega
    \end{array}
  \right]
  \label{eq:chrestenson}
\end{equation}

\begin{figure}[H]
  \centering
  \begin{subfigure}[b]{0.49\textwidth}
    \centering
    \includegraphics[width=\textwidth]{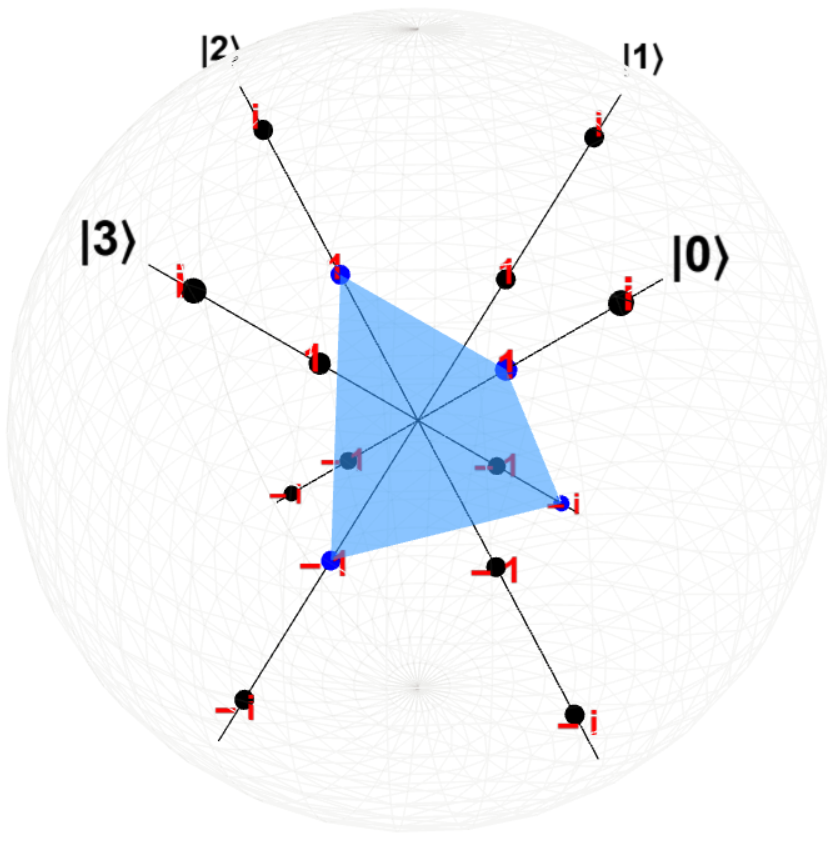}
    \caption{ $\frac{1}{2}\left( |0\rangle - |1\rangle + |2\rangle -i|3\rangle \right) = \frac{1}{2}[1,-1,1,-i]^T $\\$~$ }
  \end{subfigure}
  \hfill
  \begin{subfigure}[b]{0.49\textwidth}
    \centering
    \includegraphics[width=\textwidth]{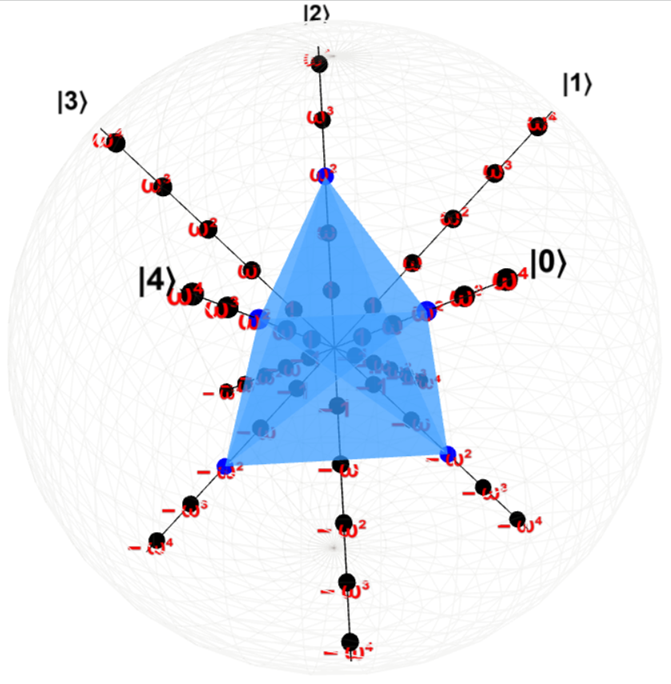}
    \caption{ $\frac{1}{\sqrt{5}}\left( \omega^2|0\rangle - \omega^2|1\rangle + \omega^2|2\rangle - \omega^2|3\rangle + \omega^2|4\rangle \right)$\\$= \frac{\omega^2}{\sqrt{5}} [1,-1,1,-1,1]^T $ }
  \end{subfigure}
  \caption{Two MAPS visualizing the mixed states of: \textbf{(a)} a ququadit ($d=4$) as the quadrilateral (4-sided polygon), and \textbf{(b)} a quintit ($d=5$) as the pentagonal (5-sided polygon), where the blue circles indicate global and relative phases, and the translucent blue polygons indicate the complete $d$-valued quantum state-space of a qudit \cite{albayaty2026maps}.}
  \label{fig:case_2}
\end{figure}

\noindent
\textbf{Example 2: Polygons of mixed states} \\
For $d \geq 3$, when there is only one blue circle visualized on the MAPS, this indicates the global phase of a pure state, as shown in Fig.~\ref{fig:global_phases_0_state}(a) and Fig.~\ref{fig:global_phases_0_state}(b). However, after choosing all relative phases (as blue circles) of $d$ mixed states for a qudit, a translucent blue $d$-sided polygon is drawn to indicate the complete quantum state-space for these mixed states. For instance, the triangle (3-sided polygon) indicates ternary mixed states for a qutrit ($d = 3$), as shown in Fig.~\ref{fig:case_1}. The quadrilateral (4-sided polygon) indicates quaternary mixed states for a ququadit ($d = 4$), as shown in Fig.~\ref{fig:case_2}(a). The pentagonal (5-sided polygon) indicates quinary mixed states for a quintit ($d = 5$), as shown in Fig.~\ref{fig:case_2}(b).

\noindent
\textbf{Example 3: Global phases of mixed states} \\
The global phase of $d$-valued mixed states for a qudit ($d \geq 3$) can be directly observed along the $|0\rangle$-axis of the MAPS, without factorizing any mathematical-based quantum representation, e.g., the Dirac or Heisenberg notation. For instance, based on the $|0\rangle$ axes for all previously shown MAPS, Fig.~\ref{fig:global_phases_0_state}(c) and Fig.~\ref{fig:global_phases_0_state}(d) visualize the global phases of 1 and $\omega^2$ for two ternary mixed states, respectively. Fig.~\ref{fig:case_1} illustrates the global phases of 1 for all three ternary mixed states. Fig.~\ref{fig:case_2}(a) and Fig.~\ref{fig:case_2}(b) demonstrate the global phases of 1 for the quaternary mixed states and $\omega^2$ for the quinary mixed states, respectively. \\

\begin{figure}[H]
  \centering
  \begin{subfigure}[b]{0.45\textwidth}
    \centering
    \includegraphics[width=\textwidth]{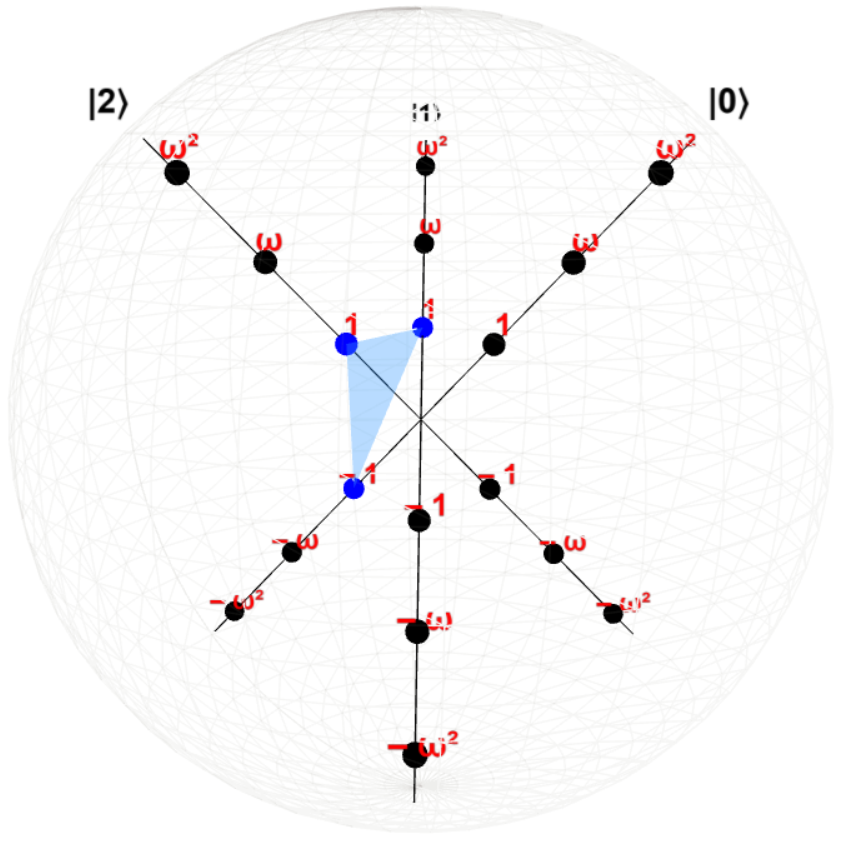}
    \caption{$CH$ and $PASS_3$ (swiveling only)}
  \end{subfigure}
  \vfill
  \begin{subfigure}[b]{0.45\textwidth}
    \centering
    \includegraphics[width=\textwidth]{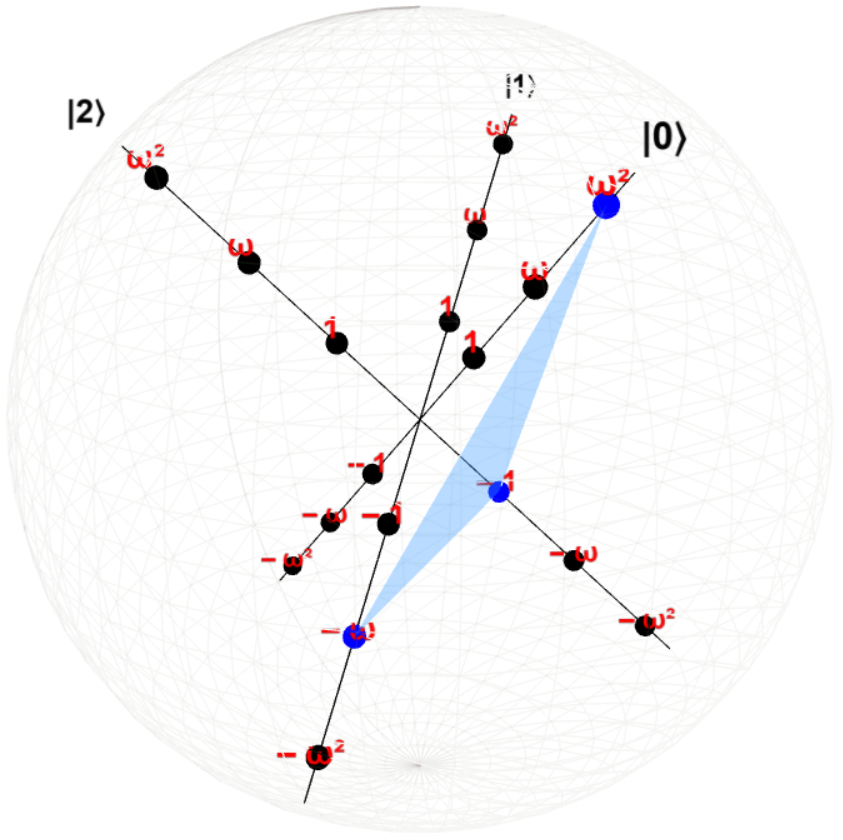}
    \caption{$CH$ and $PASS_3$ (swiveling \& shifting)}
  \end{subfigure}
  \vfill
  \begin{subfigure}[b]{0.45\textwidth}
    \centering
    \includegraphics[width=\textwidth]{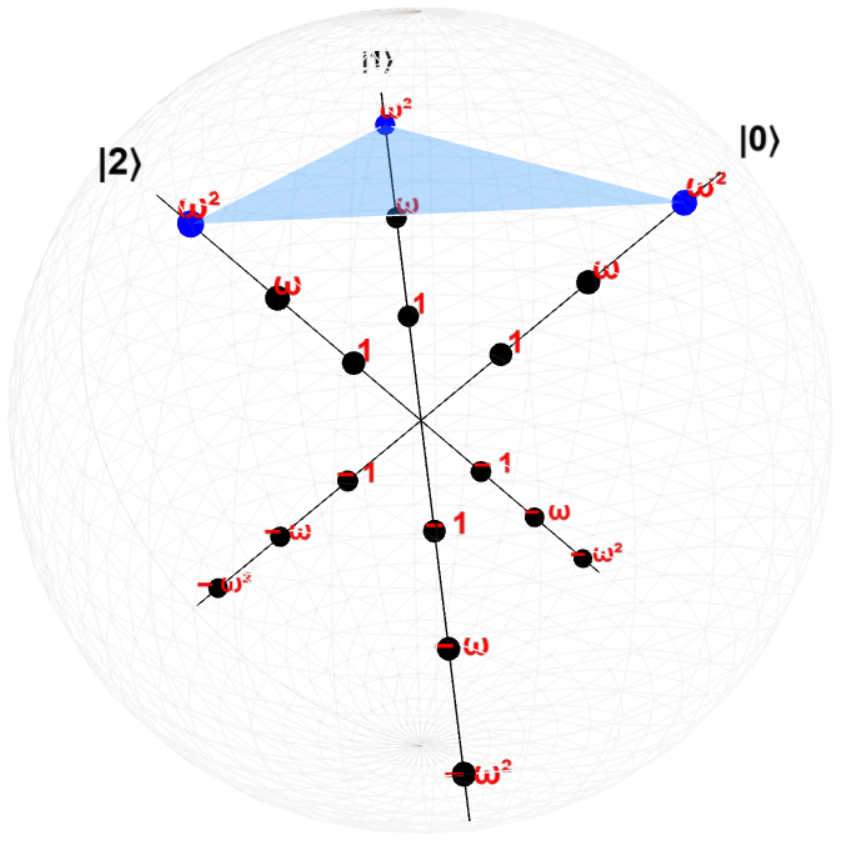}
    \caption{$CH$ and $PASS_3$ (shifting only)}
  \end{subfigure}
  \caption{Three MAPS visualizing the swiveling and shifting applied to the 3-valued mixed states of $\frac{1}{\sqrt{3}}\left[ 1,1,1 \right]$, for the outcomes of: \textbf{(a)} $\frac{-1}{\sqrt{3}} \left[ 1,-1,-1 \right]^T$, \textbf{(b)} $\frac{\omega^2}{\sqrt{3}} \left[ 1,-\omega^2,-\omega \right]^T$, and \textbf{(c)} $\frac{\omega^2}{\sqrt{3}} \left[ 1,1,1 \right]^T$, where the $|0\rangle$-axis indicates the global phases and the translucent blue triangles are the complete 3-valued quantum state-space of a qutrit \cite{albayaty2026maps}.}
  \label{fig:case_4}
\end{figure}

\noindent
\textbf{Example 4: Phase-axial swiveling and shifting gates} \\
The $PASS_d$ gate swivels and shifts the $d$-valued (pure and mixed) states of a qudit along specific $|n\rangle$ axes of the MAPS, as presented in Definition 2 above, where the $|0\rangle$-axis always indicates the global phase of the overall $d$-valued quantum state-space for a qudit ($d \geq 3$). For instance, for the $3$-valued mixed states of a qutrit shown in Fig.~\ref{fig:case_1}(a), the $PASS_3$ gate expressed in Eq.~(\ref{eq:PASS_3_1}) only swivels the $|0\rangle$ mixed state along the $|0\rangle$-axis of the MAPS, as shown in Fig.~\ref{fig:case_4}(a). Collectively, these 3-valued mixed states are swiveled from $\frac{1}{\sqrt{3}} \left[ 1,1,1 \right]^T$ with the global phase of 1 to $\frac{1}{\sqrt{3}} \left[ -1,1,1 \right]^T = \frac{-1}{\sqrt{3}} \left[ 1,-1,-1 \right]^T$ with the global phase of $-1$, as directly observed on the $|0\rangle$-axis. Notice that these 3-valued mixed states are reassigned from the top hemisphere to both the top and bottom hemispheres of the MAPS.

\begin{equation}
  PASS_3 =
  \left[
    \begin{array}{rrr}
      -1 & 0 & 0 \\
      0 & 1 & 0 \\
      0 & 0 & 1
    \end{array}
  \right]
  \label{eq:PASS_3_1}
\end{equation}

In addition, the $PASS_3$ gate expressed in Eq.~(\ref{eq:PASS_3_2}) swivels both $|1\rangle$ and $|2\rangle$ mixed states (along the $|1\rangle$ and $|2\rangle$ axes, respectively) and shifts only the $|0\rangle$ mixed state (along the $|0\rangle$-axis), as shown in Fig.~\ref{fig:case_4}(b). Collectively, these 3-valued mixed states are swiveled and shifted from $\frac{1}{\sqrt{3}} \left[ 1,1,1 \right]^T$ with the global phase of 1 to $\frac{1}{\sqrt{3}} \left[ \omega^2,-\omega,-1 \right]^T = \frac{\omega^2}{\sqrt{3}} \left[ 1,-\omega^2,-\omega \right]^T$ with the global phase of $\omega^2$, as directly observed on the $|0\rangle$-axis. Notice that these 3-valued mixed states are reassigned from the top hemisphere to both hemispheres of the MAPS.

\begin{equation}
  PASS_3 =
  \left[
    \begin{array}{rrr}
      \omega^2 & 0 & 0 \\
      0 & -\omega & 0 \\
      0 & 0 & -1
    \end{array}
  \right]
  \label{eq:PASS_3_2}
\end{equation}

Moreover, the $PASS_3$ gate stated in Eq.~(\ref{eq:PASS_3_3}) only shifts all $|0\rangle$, $|1\rangle$, and $|2\rangle$ mixed states along all $|0\rangle$, $|1\rangle$, and $|2\rangle$ axes, respectively, as shown in Fig.~\ref{fig:case_4}(c). Collectively, these 3-valued mixed states are shifted from $\frac{1}{\sqrt{3}} \left[ 1,1,1 \right]^T$ with the global phase of 1 to $\frac{1}{\sqrt{3}} \left[ \omega^2,\omega^2,\omega^2 \right]^T = \frac{\omega^2}{\sqrt{3}} \left[ 1,1,1 \right]^T$ with the global phase of $\omega^2$, as directly observed on the $|0\rangle$-axis. These 3-valued mixed states are still reside in the top hemisphere of the MAPS.

\begin{equation}
  PASS_3 =
  \left[
    \begin{array}{rrr}
      \omega^2 & 0 & 0 \\
      0 & \omega^2 & 0 \\
      0 & 0 & \omega^2
    \end{array}
  \right]
  \label{eq:PASS_3_3}
\end{equation} \\

\begin{figure}[H]
  \centering
  \begin{subfigure}[b]{0.45\textwidth}
    \centering
    \includegraphics[width=\textwidth]{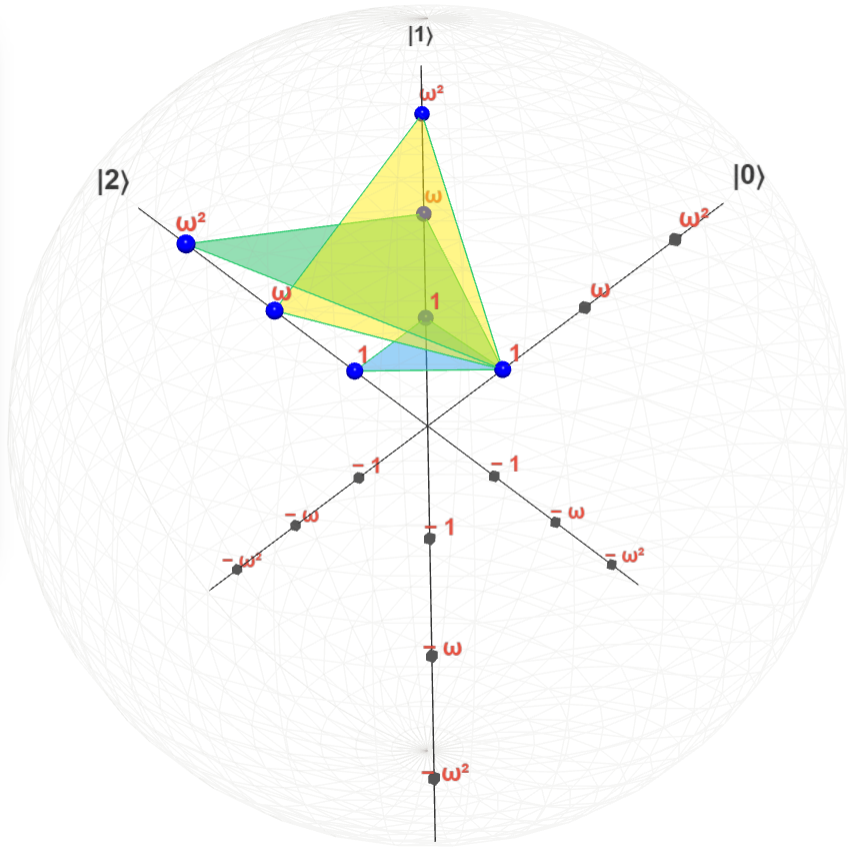}
    \caption{$CH$ only}
  \end{subfigure}
  \vfill
  \begin{subfigure}[b]{0.45\textwidth}
    \centering
    \includegraphics[width=\textwidth]{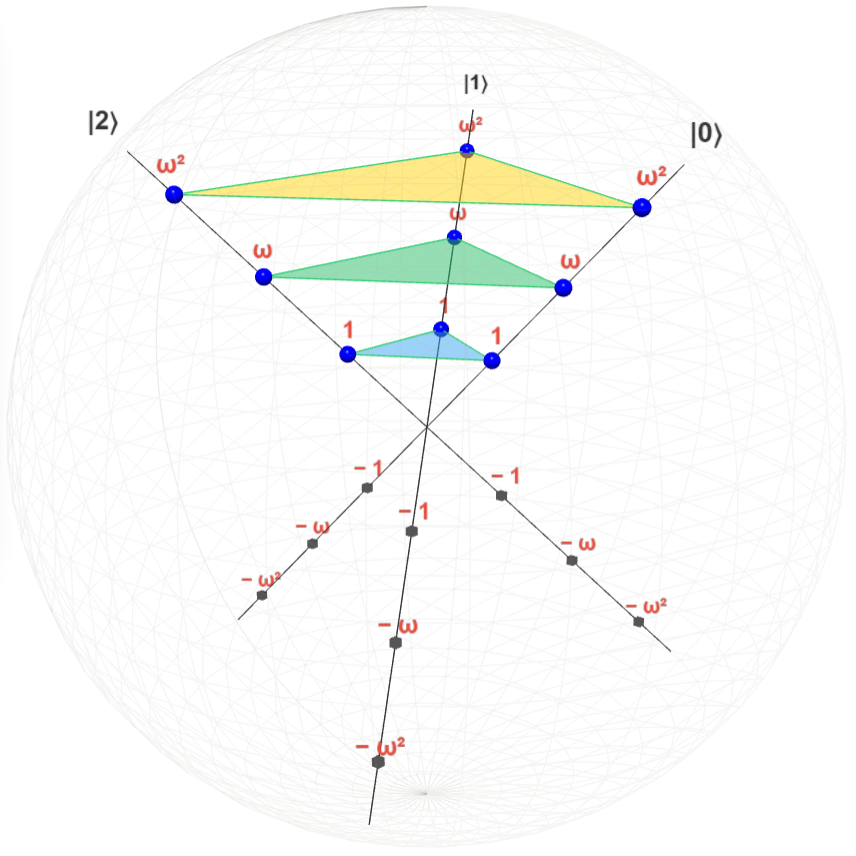}
    \caption{$CH$ and $PASS_3$ (shifting only)}
  \end{subfigure}
  \vfill
  \begin{subfigure}[b]{0.45\textwidth}
    \centering
    \includegraphics[width=\textwidth]{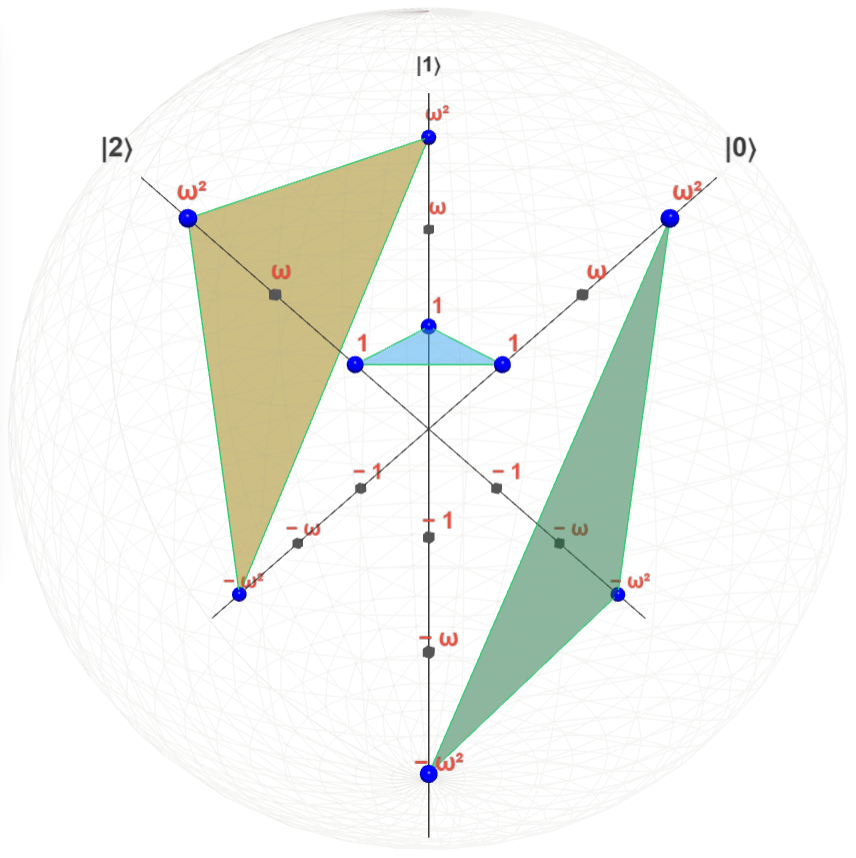}
    \caption{$CH$ and $PASS_3$ (swiveling \& shifting)}
  \end{subfigure}
  \caption{Three MAPS visualizing a set of $CH$ and $PASS_3$ gates applied to three qutrits, resulting in three different 3-valued mixed states. Every $|0\rangle$-axis for every qutrit indicates its global phase, and the translucent blue, green, and yellow triangles are the 3-valued mixed states for the first, second, and third qutrits, respectively \cite{albayaty2026maps}.}
  \label{fig:case_5}
\end{figure}

\noindent
\textbf{Example 5: Geometrical comparison of diverse mixed states} \\
The MAPS, as a single $S^2$ framework, can also be utilized for geometrically comparing diverse $d$-valued quantum (pure and mixed) states for a set of qudits, where $d \geq 3$. For instance, Fig.~\ref{fig:case_5}(a) illustrates three different 3-valued mixed states after applying the $CH$ gate stated in Eq.~(\ref{eq:chrestenson}) to three qutrits of initial pure states $|0\rangle$, $|1\rangle$, and $|2\rangle$, respectively.

In addition, Fig.~\ref{fig:case_5}(b) demonstrates three different 3-valued mixed states of three qutrits. Initially, the $CH$ gate stated in Eq.~(\ref{eq:chrestenson}) is applied to all three qutrits of initial pure states $|0\rangle$, $|1\rangle$, and $|2\rangle$, respectively. Then, the $PASS_3$ gate expressed in Eq.~(\ref{eq:PASS_3_4}) is only applied to the second qutrit. Finally, the $PASS_3$ gate stated in Eq.~(\ref{eq:PASS_3_5}) is only applied to the third qutrit. Notice that these two $PASS_3$ gates only perform the shifting quantum operation without any swiveling, in the same top hemisphere of the MAPS.

\begin{equation}
  PASS_3 =
  \left[
    \begin{array}{rrr}
      \omega & 0 & 0 \\
      0 & 1 & 0 \\
      0 & 0 & \omega^2
    \end{array}
  \right]
  \label{eq:PASS_3_4}
\end{equation}

\begin{equation}
  PASS_3 =
  \left[
    \begin{array}{rrr}
      \omega^2 & 0 & 0 \\
      0 & 1 & 0 \\
      0 & 0 & \omega
    \end{array}
  \right]
  \label{eq:PASS_3_5}
\end{equation}

Moreover, Fig.~\ref{fig:case_5}(c) illustrates three different 3-valued mixed states of three qutrits. Initially, the $CH$ gate stated in Eq.~(\ref{eq:chrestenson}) is applied to all three qutrits of initial pure states $|0\rangle$, $|1\rangle$, and $|2\rangle$, respectively. Then, the $PASS_3$ gate expressed in Eq.~(\ref{eq:PASS_3_6}) is only applied to the second qutrit. Finally, the $PASS_3$ gate stated in Eq.~(\ref{eq:PASS_3_7}) is only applied to the third qutrit. Notice that these two $PASS_3$ gates perform both swiveling and shifting quantum operations, in both top and bottom hemispheres of the MAPS.

\begin{equation}
  PASS_3 =
  \left[
    \begin{array}{rrr}
      \omega^2 & 0 & 0 \\
      0 & -\omega & 0 \\
      0 & 0 & -1
    \end{array}
  \right]
  \label{eq:PASS_3_6}
\end{equation}

\begin{equation}
  PASS_3 =
  \left[
    \begin{array}{rrr}
      -\omega^2 & 0 & 0 \\
      0 & 1 & 0 \\
      0 & 0 & \omega
    \end{array}
  \right]
  \label{eq:PASS_3_7}
\end{equation}

Finally, the MAPS, as an $S^2$ framework, can be practically integrated with various $d$-valued quantum circuit simulators, such as the QuDiet \cite{chatterjee2023qudiet}, GCAMPS \cite{harper2026gcamps}, as well as our ternary and hybrid (qubits + qutrits) circuit simulators \cite{albayaty2025simulators}, to effectively visualize and geometrically analyze the quantum characteristics of different $d$-valued quantum (pure and mixed) states for a set of qudits, where $d \geq 3$.

\section{Conclusions}
In quantum computing and information science, the Bloch sphere, as a three-dimensional $S^2$ geometry, plays an important role in visualizing the 2-valued pure (basis) and mixed (superimposed) states of a qubit. However, the global and relative (local) phases of these states are not directly visualized along the three intersecting spatial axes of the Bloch sphere. Hence, additional mathematical formalisms are required to describe the complete 2-valued quantum states of a qubit. Moreover, the Bloch sphere cannot visualize higher $d$-valued quantum states of a qudit ($d \geq 3$), such as a qutrit, ququadit, and quintit for $d$ = 3, 4, and 5, respectively. Therefore, complicated geometrical and topological representations are necessary to visualize the complete $d$-valued quantum state-space of a qudit.

For these two reasons, this paper introduces and formalizes a new generalized three-dimensional $S^2$ framework to visualize and describe the higher $d$-valued quantum states of a qudit, in terms of ease of illustration, structural simplicity, and natural representation, without requiring any additional mathematical formalism and complicated topological structure. We called this new visualization $S^2$ framework the ``multi-axial projective sphere (MAPS)''. In general, the MAPS is geometrically constructed using $n$ projectional intersecting spatial axes. Every spatial axis is mapped to one $d$-valued quantum state of a qudit with its corresponding set of relative phases $\pm \omega_d^k$, and the $|0\rangle$-axis of the MAPS always indicates the global phase for a qudit, where $d \geq 3$, $0 \leq n \leq d-1$, and $k \in \mathbb{Z}$. In our research, we also propose a group of novel ``$d$-valued phase-axial swiveling and shifting gate $PASS_d$'', to swivel (rotate) and shift (scale) the $d$-valued quantum states of a qudit along these $n$ spatial axes. This group consists of $|\pm \omega_d^k |^d-1$ permutative $PASS_d$ gates. The MAPS and these permutative $PASS_d$ gates preserve the quantum properties, visual distinguishability, and entanglement structures for all $d$-valued quantum states of a qudit.

Our future work will focus on employing the MAPS and $PASS_d$ gates to visually build and cost-effectively construct various $d$-valued quantum operators, e.g., arithmetic circuits, counters, and comparators, without any $d^{\otimes m} \times d^{\otimes m}$ matrices multiplication, where $m \geq 1$, which is the total number of qudits for a quantum operator. In addition, the MAPS and $PASS_d$ gates can also be integrated with various $d$-valued quantum circuit simulators, to effectively visualize and geometrically analyze a set of $d$-valued quantum states of $m$ qudits. Finally, as a generalized $S^2$ framework, the MAPS could be used for visualizing high-dimensional data for scientific and engineering applications, such as machine learning, quantum machine learning, quantum chemistry, just to name a few, where every spatial axis of the MAPS represents a single feature (dimensionality) of such data with its corresponding distinct (numerical or textual) values.

\section*{Data Availability}
All relevant data are available upon request from the corresponding author (A.A.-B.).

\printbibliography




\end{document}